\documentclass[11pt,a4paper]{article}
\usepackage[T1]{fontenc}
\usepackage[utf8]{inputenc}
\usepackage[margin=2.5cm]{geometry}
\usepackage{amsmath,amssymb,bm}
\usepackage{booktabs}
\usepackage{enumitem}
\usepackage{graphicx}
\usepackage{tabularx,array,etoolbox}
\graphicspath{{figures/}}
\usepackage[hidelinks]{hyperref}
\usepackage[numbers,square,sort&compress]{natbib}
\usepackage{titlesec}

\usepackage{longtable,xurl}

\titleformat{\section}{\large\bfseries}{\thesection.}{0.5em}{}
\titleformat{\subsection}{\normalsize\bfseries}{\thesubsection}{0.5em}{}
\setlist{itemsep=2pt,topsep=3pt}

\newcommand{\E}{\mathbb{E}}
\newcommand{\R}{\mathbb{R}}

\title{\textbf{From Vector Autoregressions to AI-based Time Series Forecasting: A Review}}

\author{Likai Chen\thanks{Washington University in St.\ Louis. \texttt{likai.chen@wustl.edu}.}
\and Weining Wang\thanks{University of Bristol. \texttt{weining.wang@bristol.ac.uk}.}}

\date{14 November 2026}
\begin{document}
\maketitle

\begin{abstract}
Forecasting is a central goal of time-series analysis. This review examines
three developments in AI-based forecasting: transformer architectures,
pretrained foundation models for zero-shot forecasting, and diffusion-based
generative forecasters. We connect these methods to the econometric tradition
built around the vector autoregression (VAR) by examining how each models
the conditional distribution of future observations given the past or
predicts selected features of that distribution. The review is organized
around three long-standing challenges: \emph{high dimensionality},
\emph{nonstationarity}, and \emph{nonlinearity}. We argue that modern
methods expand the classical forecasting template by allowing more flexible
dynamics, drawing on larger training corpora, and representing richer
predictive distributions. Yet the well-established tools for hypothesis
testing and structural analysis in classical econometrics do not carry over
directly to many modern forecasting models. We close by asking when these
modelling choices yield reliable forecasting gains under dependence and
structural change.

\end{abstract}
\section{Introduction}
\label{sec:intro}
Forecasting is central to applied econometrics, with uses ranging from monetary
policy analysis and risk management to demand planning and nowcasting. For four
decades, the vector autoregression (VAR) of \citet{sims1980macroeconomics} has
provided a coherent way to study how multiple time series move together. Its
appeal comes not only from forecasting performance, but also from the
interpretive apparatus built around it: impulse-response analysis, Granger
causality \citep{granger1969investigating}, and the cointegration and
error-correction framework of \citet{engle1987cointegration} and
\citet{johansen1991estimation}. Standard graduate treatments
\citep{lutkepohl2005new,hamilton1994time} still organize the field around this
core. For many problems, especially small systems and settings where
interpretation, valid inference, or long-run relationships matter, the VAR
remains the natural starting point.

At the same time, methods imported from machine learning have expanded what is
possible at scale. The transformer architecture \citep{vaswani2017attention} is
a leading example. {For time-series forecasting, attention combines
transformed lagged observations using weights that depend on the observed window.}
Foundation models \citep{ansari2024chronos,das2024timesfm,woo2024moirai}
are pretrained on large and diverse collections of time series. In zero-shot
forecasting, they use the observed history of a new series to produce forecasts
without re-estimating their parameters. Denoising diffusion models
\citep{ho2020denoising,rasul2021timegrad} add a generative layer, sampling
future paths from the predictive distribution and thereby representing
forecast uncertainty more directly. These developments concern complementary
parts of a forecasting specification: transformers provide an architecture,
pretraining supports reuse across forecasting tasks, and diffusion provides
a generative forecasting mechanism. They can therefore be combined within the
same model. Although these developments are often presented as a break from
classical statistics, we argue that they are better understood as new
approaches to the same forecasting problem.

\par\begingroup\ignorespaces 

Earlier extensions of the basic VAR and other forecasting models
already allowed for nonlinear dynamics and probabilistic forecasting.
Nonlinear econometric models allow dynamics to depend on the
state of the system, while time-varying parameter models allow
coefficients to change over time. State-space methods describe
observations through unobserved states that evolve over time.
Earlier probabilistic neural forecasters, such as DeepAR
\citep{salinas2020deepar}, already learned across related series
and produced predictive distributions. Our focus is on
input-dependent weighting through attention, pretraining across
diverse collections of series and zero-shot transfer, and
diffusion models for generating joint future paths.
\par\endgroup 

\par\begingroup\ignorespaces 
To compare these approaches, we focus on the conditional law of
future observations, or a specified functional of that law, given the
available history and admissible covariates. A linear VAR specifies a conditional mean; its predictive law also
requires an innovation distribution. Gaussian innovations yield a joint Gaussian
future path conditional on fixed parameters and a suitable Gaussian initial-state
specification. A transformer may output means, quantiles, a density, or samples,
depending on the forecasting layer and loss. Diffusion uses a learned sampling
mechanism to represent a flexible implicit conditional family.
Predictive flexibility does not, however, by itself identify an economic shock
or justify uncertainty statements. Nonlinear models can nevertheless support
impulse responses and predictive-relevance restrictions
\citep{koop1996impulse,tank2022neural}. We therefore distinguish defining
an estimand, identifying it under stated assumptions, and conducting
valid inference.
\par\endgroup 

To illustrate these distinctions, consider a small macroeconomic panel containing inflation,
output growth, and a short-term interest rate. A classical VAR uses their joint
history to forecast future values. An identified Structural VAR asks how a monetary policy
shock propagates through inflation and output. {A vector error-correction model (VECM) is relevant when the selected variables are integrated and cointegrated. {An attention model can assign different weights to past dates
depending on the recent paths of inflation, output growth, and interest rates.}} A diffusion forecaster describes
uncertainty by sampling possible future macroeconomic paths. We return to this example throughout to distinguish forecasting performance
from the econometric questions of identification, inference, and
nonstationarity. {Appendix~\ref{sec:application} turns this
example into a reproducible public-data forecasting comparison, with
saved forecasts and executable code.}

\par\begingroup\ignorespaces 
We compare these forecasting approaches through three distinct challenges.
A method may address one without resolving the others.
\begin{enumerate}
\item \textbf{High dimensionality.}
A VAR with $p$ variables and $L$ lags contains $p^2L$
autoregressive coefficients. Reliable estimation with limited
data requires restrictions on effective complexity. LASSO-type
methods exploit exact or approximate sparsity, whereas Bayesian
VARs control complexity through prior shrinkage
\citep{basu2015,kock2015,banbura2010large}.
Factor models instead reduce dimensionality by representing
common variation through a low-rank component
\citep{stock2002forecasting,baing2002}.
Transformers offer a different way of structuring the forecasting
function. Rather than requiring sparse autoregressive
coefficients or a low-rank common factor representation of the
observed system, attention constructs input-dependent
interactions through shared learned parameters
\citep{vaswani2017attention}. The resulting attention weights
adapt to the observed window even when the fitted parameters
remain fixed. 
{Diffusion models contribute to joint probabilistic
forecasting by learning denoising maps that generate future
paths \citep{rasul2021timegrad,tashiro2021csdi}.
Representing a $pH$-dimensional future path, where $H$ denotes
the forecast horizon, does not itself reduce the statistical
complexity of estimating its distribution. The diffusion
mechanism provides neither automatic dimension reduction nor
a general solution to high-dimensional estimation.
Its effective complexity depends on the denoising architecture,
learned representations, and any additional regularization.
Under suitable assumptions, diffusion models can nevertheless
adapt to underlying low-dimensional structure, as discussed
in Section~\ref{sec:diff-evaluation}.}
Cross-sectional dimension, context length, parameter count,
and forecast dimension therefore remain distinct considerations.

\item \textbf{Nonstationarity.}
Classical time-series methods address structural breaks,
gradual parameter drift, and stochastic trends through
change-point models, time-varying parameters, and unit-root
and cointegration analysis
\citep{bai1998breaks,primiceri2005tvp,johansen1991estimation}.
Threshold and switching models describe state-dependent
dynamics, which can also arise in stationary processes
\citep{tsay1998threshold,krolzig1997markov}.
Transformers offer a flexible form of state-dependent
information aggregation: attention weights adjust to the
observed history even when the fitted network parameters
remain fixed \citep{vaswani2017attention}.
Conditional diffusion models similarly use the observed
history to shape the predictive distribution
\citep{rasul2021timegrad,tashiro2021csdi}.
Such conditioning, like input-dependent attention, does not
by itself specify how the underlying process changes over time.
Both approaches can nevertheless complement explicit models
of structural instability and long-run relationships.
A natural direction is therefore to combine these forecasting
architectures with change-point detection and model updating,
time-varying parameters, or error-correction mechanisms,
linking flexible predictive modelling to the relevant
time-series structure.
\item \textbf{Nonlinearity.}
Threshold models, switching models, trees, kernels, and neural
networks already permit nonlinear or state-dependent prediction
\citep{tsay1998threshold,krolzig1997markov,breiman2001forests,
friedman2001boosting,masini2023ml}.
Within the neural-network class, transformers learn predictive
relationships through input-dependent attention and nonlinear
feed-forward transformations \citep{vaswani2017attention}.
Tokenization determines whether interactions are constructed
across time points, temporal patches, or variable histories
\citep{nie2023patchtst,liu2024itransformer}.
Diffusion models provide a complementary representation of
the conditional predictive distribution, using learned
nonlinear denoising maps to generate future paths
\citep{rasul2021timegrad,tashiro2021csdi}.
Attention can also form part of the denoising network, combining
learned interactions with generative forecasting.
The resulting predictive output depends on the architecture,
conditioning information, and training objective.
\end{enumerate}
\par\endgroup 

The remainder of the review develops this comparison in detail. Section~\ref{sec:classical}
reviews the classical foundations and locates the three challenges within them.
{Section~\ref{sec:transformers} compares how VARs and attention
use lagged observations.} In particular,
Subsection~\ref{sec:tokenize} focuses on the representational
question inherited from language modelling: what counts as a token of a time
series? Section~\ref{sec:fm} reviews foundation models and zero-shot
forecasting, emphasizing the shift from per-data-set estimation to cross-corpus
pretraining. Section~\ref{sec:diffusion} reviews diffusion-based generative
forecasters, emphasizing how learned denoising mechanisms represent and sample joint predictive distributions.
Section~\ref{sec:synthesis} offers a synthesis and summary.
{Appendix~\ref{sec:practice} develops a practical implementation workflow.}
{Appendix~\ref{sec:application} applies it to the running macroeconomic example.}
{Appendix~\ref{app:gifteval} documents the GIFT-Eval comparison.}
Throughout, we aim
less at exhaustive coverage than at a coherent narrative connecting modern
forecasting methods to econometric concerns. Recent surveys provide broader
coverage of transformer methods \citep{wen2023transformers} and diffusion
models \citep{yang2024survey,meijer2024rise}.

Throughout, $\bm{y}_t\in\R^{p}$ denotes the vector of $p$ series observed at date $t$, and $T$
denotes the sample length. The symbol $L$ denotes the length of the look-back window. In
Section~\ref{sec:classical}, this is the lag order of the vector autoregression. From
Section~\ref{sec:transformers} onward, it is the context length of the sequence model. In both cases,
the input is a window of the $L$ most recent observations. The symbol $H$ denotes the forecast
{horizon and $h$ indexes individual forecast steps. In the attention walkthrough, the input ends at $t-1$ and the targets
are $\bm y_{t:t+H-1}$. In the diffusion section, the forecast origin
is $t$ and the target is the future path $\bm y_{t+1:t+H}$.
Appendix~\ref{sec:application} uses the same forecast origin and
evaluates point forecasts of $\bm y_{t+h}$ at $h\in\{1,3,6,12\}$. The embedding map is denoted by
$\varphi$ and $\mathbb E$ is reserved for expectation.}

\section{Classical multivariate foundations}
\label{sec:classical}

For a $p\times 1$ vector $\bm{y}_t$, the vector autoregression of order $L$ is
\begin{equation}
\bm{y}_t = \bm{\nu} + \sum_{i=1}^{L} A_i \bm{y}_{t-i} + \bm{\varepsilon}_t,
\qquad \bm{\varepsilon}_t \sim \mathrm{iid}\,(\bm{0}, W),
\label{eq:var}
\end{equation}
where $A_i \in \R^{p\times p}$ are coefficient matrices and $W$ is the innovation covariance matrix,
{which is generally non-diagonal.}
In the macroeconomic panel example, $\bm y_t$ might collect inflation, output
growth, and the policy rate. The coefficient matrix $A_i$ records how the
$i$th lag of each variable enters each equation. Lagged inflation may help
forecast the policy rate, lagged policy rates may help forecast output growth,
and so on. This is a reduced-form predictive object.

When the roots
of $\det(I_p - A_1 z - \cdots - A_L z^L)$ lie outside the unit circle, the process is stable and
admits the Wold representation
\[
\bm{y}_t = \bm{\mu} + \sum_{h\ge 0} B_h \bm{\varepsilon}_{t-h},
\]
with $B_0 = I_p$. Augmenting \eqref{eq:var} with moving-average terms gives the vector
autoregressive moving-average (VARMA) class. Under the corresponding autoregressive stability condition,
a VARMA model also admits an infinite moving-average representation. It can
therefore represent current outcomes as weighted accumulations of past shocks,
often with fewer parameters than a finite VAR, although estimation and
identification are harder \citep{lutkepohl2005new}. 
\subsection{Forecasting, impulse responses, and Granger causality}
\label{sec:var-uses}

The conditional mean in \eqref{eq:var} provides the one-step forecast
given the lagged observations. Multi-step forecasts are obtained
by iterating the system.
Beyond producing forecasts, a fitted VAR is used to answer two
further questions, both directly related to what modern methods
do and do not deliver.
The first is \emph{propagation}: what happens after an innovation
or shock enters the system? This leads to impulse-response analysis.
The second is \emph{predictive relevance}: does the history of one
series improve forecasts of another, given the remaining information?
This leads to Granger-causality testing.
Both questions arise in linear and nonlinear models.
A structural interpretation requires a clearly defined shock
and identifying assumptions, while formal tests of predictive
relevance require an appropriate reference distribution
under the null.

For propagation, consider first the linear VAR in \eqref{eq:var}.
Propagation is determined by the lag coefficient matrices.
The $(i,j)$ entry of $B_h$ measures the response of variable $i$
at date $t+h$ to a one-unit increase in $\varepsilon_{j,t}$,
with all other innovations held fixed. Plotting these entries
against the horizon $h$ gives an impulse-response function.
However, $W$ is generally not diagonal. Hence errors are
correlated across equations and an innovation to one variable
alone is not yet a well-defined structural experiment.
To interpret an impulse response as the effect of an economically
meaningful shock, such as a monetary policy shock, the shock of interest must be identified from the reduced-form innovations. This is the identification problem. Standard solutions
impose additional assumptions, such as a recursive ordering of
contemporaneous effects or restrictions on the signs or long-run
impacts of particular shocks \citep{blanchard1989,kilian2017}.

In the running macroeconomic example, an impulse response might
trace the effect of a monetary policy shock on output growth and
inflation over subsequent periods. Interpreting an innovation in
the equation for the policy rate as a monetary policy shock
requires additional identifying assumptions.
The same propagation question can be studied in nonlinear VARs.
Generalized impulse responses compare conditional forecasts
following a specified shock with a reference forecast, allowing
the response to depend on the observed history and on the size
and sign of the shock \citep{koop1996impulse}.

\par\begingroup\ignorespaces

Turning to predictive relevance, the question of
\citet{granger1969investigating} can be formulated without a
constant-coefficient model. Let $\mathcal F_{t-1}$ denote the
specified information available before date $t$, and let
$\mathcal F_{t-1}^{(-j)}$ remove the history of series $j$ while
retaining the other specified information. For an integrable
outcome $y_{i,t}$, one-step conditional-mean noncausality is
the restriction
\begin{equation}
\E[y_{i,t}\mid\mathcal F_{t-1}]
=
\E[y_{i,t}\mid\mathcal F_{t-1}^{(-j)}]
\quad\text{almost surely}.
\label{eq:mean-noncausality}
\end{equation}
This is a restriction on a conditional-mean function.
Noncausality for the full predictive law instead requires
equality of the corresponding conditional distributions.
Equal means alone do not establish that stronger property.

Under a correctly specified linear VAR with nondegenerate
regressors, the mean restriction reduces to zero coefficients
on all included lags of series $j$ in the equation for series $i$.
{Standard Wald tests jointly test these restrictions.}
Their validity depends on assumptions that justify
{the reference distribution of the test statistic
under the null.} In a nonlinear VAR, the corresponding restriction is that the
conditional-mean forecasting function does not depend on the
history of the excluded series \citep{tank2022neural}.
Predictive relevance should not be interpreted as the effect
of a policy intervention. The distinction is therefore between
defining an estimand, identifying it from the observed data
distribution under stated assumptions, and conducting valid
inference for it.

\par\endgroup 

\subsection{Unit roots, cointegration, and error correction}
\label{sec:cointegration}

The stability condition stated above is a restriction on the VAR coefficients,
rather than part of the VAR definition. A constant-coefficient linear VAR may
also contain unit roots and cointegrating relations. Many economic series are
integrated of order one. Unit-root tests \citep{dickey1979,phillips1988}
help assess this possibility. Regressions among unrelated integrated series
may be spurious \citep{GRANGER1974111}.

Cointegration provides a way to retain long-run relationships: $I(1)$ series
may share a stationary long-run combination. In this case, the VAR can
be written in the vector error-correction form, with deterministic terms
suppressed here,
\begin{equation}
\Delta \bm{y}_t = \bm{\alpha}\,\bm{\beta}^{\top}\bm{y}_{t-1}
+ \sum_{j=1}^{L-1} \Gamma_j \,\Delta\bm{y}_{t-j} + \bm{\varepsilon}_t,
\label{eq:vecm}
\end{equation}
where $\Delta$ denotes the first difference, $\bm{\beta}\in\R^{p\times r}$
collects the $r$ cointegrating vectors defining the stationary combinations
$\bm{\beta}^{\top}\bm{y}_t$, $\bm{\alpha}\in\R^{p\times r}$ gives the speeds at
which deviations from those combinations are corrected, and the $\Gamma_j$ are
short-run dynamic coefficients. The cointegrating rank $r$ and the coefficient
matrices can be estimated by the reduced-rank procedure of
\citet{johansen1991estimation,johansen1995}, building on
\citet{engle1987cointegration} and the unit-root distribution theory of
\citet{dickey1979,phillips1988}.

{If the data are cointegrated, a finite-lag model in
differences that omits relevant level information generally
excludes the equilibrium-restoring term in \eqref{eq:vecm}.
Neural and generative forecasters face the same risk when their
inputs or preprocessing discard the information needed to
represent such adjustment. The absence of explicit cointegrating
restrictions does not, however, imply that a flexible model
supplied with suitable inputs is incapable of learning long-run
adjustment.}
In the macroeconomic example, cointegration would matter if some
variables shared a stable long-run relation, so that deviations from that
relation predict later adjustment. A model estimated only in differences may nevertheless forecast short-run movements well.

\par\begingroup\ignorespaces 
\subsection{Beyond the basic VAR}
\label{sec:intermediate}

Established extensions and alternative forecasting approaches address high
dimensionality, parameter instability, and nonlinearity through different
modelling restrictions.
\par\endgroup High dimensionality is already binding in the VAR, since the unrestricted
system \eqref{eq:var} contains $O(p^2L)$ parameters.
{The main remedies control effective complexity through stated restrictions.}\par\begingroup\ignorespaces 

LASSO-type methods address high-dimensional estimation by shrinking
and selecting coefficients, including the coefficients on the
lagged values of each variable in large VARs
\citep{tibshirani1996lasso,basu2015,kock2015}.
Their statistical justification relies on exact or approximate
sparsity: only a small subset of coefficients is nonzero or
sufficiently large. Suitable conditions on temporal dependence
and the regressors are also needed. Penalty strength and lag
order must be selected using training and validation information,
with data requirements governed by effective sparsity and
dependence. Elastic net combines sparsity with additional
shrinkage \citep{zou2005elastic}.
For regression systems with temporal and cross-sectional
dependence, \citet{chernozhukov2021lasso} develop joint penalty
selection and debiased simultaneous inference. Related
developments include regularized and debiased estimation
of spatial panel networks \citep{chernozhukov2026spatial}
and LASSO-based construction of informative instruments
in dynamic panel models \citep{chernozhukov2024arellano}.

Bayesian VARs address estimation uncertainty in large systems by shrinking
coefficients towards a chosen prior specification, including priors of the Minnesota family
\citep{litterman1986forecasting,banbura2010large}. Their performance depends
on the prior centre and tightness, including the relative shrinkage of
own lags, cross-variable lags, and more distant lags. Conventional
continuous shrinkage priors need not set coefficients exactly to zero.
The relevant restriction is therefore the prior structure and its strength
relative to the model dimension and available sample.

Dynamic factor models summarize a large panel through a small number of latent
common components \citep{stock2002forecasting,baing2002}. This dimension
reduction relies on a low-rank common component, together
with sufficiently strong factors and restrictions on the remaining
idiosyncratic dependence. The low-rank restriction concerns the common
component, not necessarily the full data or covariance matrix. It is
distinct from assuming that most observable VAR coefficients are zero.

\par\endgroup 

In a VAR, abrupt instability can be modelled by allowing the coefficient
matrices in \eqref{eq:var} to change at unknown dates. Estimation then concerns
the number and locations of the breaks as well as the coefficients within
each segment.
{Change-point methods rely on assumptions about break magnitudes and the
information available within each segment \citep{bai1998breaks}.}
For univariate mean shifts with independent Gaussian noise,
under suitable spacing and jump-size conditions, segmentation
estimators can be consistent with near-optimal localization
rates even when the number of breaks grows with the sample.
Wild binary segmentation is one example \citep{fryzlewicz2014wbs}.
 More recently, neural networks have also been trained as change-point
detectors with accompanying theory (see, for example,
\citep{li2024automatic}).

For gradual instability, parameters may instead be allowed
to drift smoothly. This is the idea behind the locally stationary framework of
\citet{dahlhaus1997fitting}, where the process is approximately stationary within short windows and
can be estimated by kernel or local-likelihood methods. {Time-varying-parameter VARs in econometrics likewise specify how
coefficients evolve, with the evolution law and its regularization
controlling the extent of adaptation \citep{primiceri2005tvp}. These
specifications address forms of parameter instability, rather than
nonlinearity or high dimensionality alone.}

\par\begingroup\ignorespaces 

The VAR in \eqref{eq:var} has a linear conditional mean with constant
coefficients. Nonlinear and state-dependent extensions relax these
restrictions.
Linearity of the conditional mean
does not itself require Gaussian innovations.
Threshold models represent nonlinear dynamics through regimes determined
by an observed threshold variable, whereas Markov-switching models use
latent regimes \citep{tsay1998threshold,krolzig1997markov}. These models
can be stationary or nonstationary, depending on their parameters,
regime dynamics, and other assumptions. Nonlinearity or regime dependence
alone does not determine stationarity. Threshold error-correction models
combine nonlinear adjustment with cointegrating restrictions
\citep{hansen2002threshold}. Estimation requires sufficient information
within regimes, and large regime-specific systems may still need
regularization.

Random forests and gradient boosting learn nonlinearities and
interactions through trees \citep{breiman2001forests,friedman2001boosting,
januschowski2022trees}. Kernel methods and neural networks provide other
nonlinear forecasting classes \citep{masini2023ml}. Their usefulness
depends on the features, model complexity, regularization, and tuning
relative to the available training information \citep{goulet2022ml}.
\par\endgroup 
\par\begingroup\ignorespaces 

These extensions also allow the questions of propagation and predictive
relevance in Section~\ref{sec:var-uses} to be studied beyond the basic VAR.
{In nonlinear or time-varying models, impulse responses can depend on
the history, the date, and the size or sign of the shock.}\citet{koop1996impulse} develop generalized impulse responses for nonlinear
as well as linear multivariate models. Applying this
idea to a forecasting model requires an explicit shock or innovation, a
reference comparison, and a law describing subsequent propagation. Time-varying models likewise
require their conditioning state and parameter-evolution specification to
be made explicit \citep{primiceri2005tvp}.

\citet{tank2022neural} provide a nonlinear example using componentwise
multilayer perceptrons or recurrent networks with structured sparsity
penalties to exclude particular input histories. This illustrates how neural
architecture and regularization can encode predictive-relevance restrictions.
A selected exclusion in a fitted network is not, by itself, a population
noncausality result or a calibrated hypothesis test. The inferential target,
conditioning information, approximation and estimation error, temporal
dependence, and model selection all matter. In particular, attention weights
alone are not tests of Granger noncausality.

\par\endgroup 

\par\begingroup\ignorespaces 

The transformer forecasters considered in this review belong to the
nonlinear machine-learning class described above: a transformer is a particular
neural-network architecture, not a separate alternative to neural
networks. Its distinctive structure combines token embeddings, shared
attention projections, and input-dependent interactions, with parallel
computation across observed tokens \citep{vaswani2017attention}.
Originally developed for language sequence modelling, the architecture
can also accommodate multiple data modalities: for example, text tokens
and image patches can be embedded and related through attention in
vision--language models \citep{kim2021vilt}. Time-series transformers
adapt this token-based structure to observations, temporal patches, or
variable histories. These architectural choices complement, rather than
replace, explicit regularization. Section~\ref{sec:transformer-adaptations} discusses how LASSO-type
penalties can be applied to selected trainable network weights.

\par\endgroup 

\par\begingroup\ignorespaces 

{For a fitted transformer, a structural interpretation requires a clearly
defined shock and identifying assumptions.}
Fitted network parameters may remain fixed while
attention weights change with the input window. Such context dependence does
not eliminate inferential targets. It changes how they are represented and
what estimation and uncertainty procedures must be justified.
Figure~\ref{fig:var2attn} illustrates the predictive parameterizations and does
not imply that statistical inference is possible for VARs but impossible for
transformers.
Table~\ref{tab:bridge} compares the restrictions, data requirements,
covariate support, forecast outputs, and computational costs of these
forecasting specifications.

\par\endgroup

\section{The transformer era}
\label{sec:transformers}

\par\begingroup\ignorespaces 
Building on Section~\ref{sec:intermediate}, we now examine
transformers as a specific neural-network architecture for
forecasting. Earlier neural forecasters include recurrent
models such as DeepAR \citep{salinas2020deepar} and
basis-expansion or hierarchical architectures such as
N-BEATS and N-HiTS
\citep{oreshkin2020nbeats,challu2023nhits}.
Our focus here is on self-attention and the construction
of tokens. {We first introduce the attention mechanism.} We then
explain how the resulting representations produce forecasts, compare
time-point, patch, variable, and quantized tokenization, and review
architectural adaptations and forecasting evidence.
\par\endgroup 

\subsection{The basic transformer mechanism}
\label{sec:attn-mechanism}

The transformer of \citet{vaswani2017attention} replaces recurrence with
self-attention. A token is the basic input unit processed by the model,
and an embedding maps it to a vector. For the time-point representation
used to fix ideas here, each token represents one observed multivariate
vector. Section~\ref{sec:tokenize} compares alternative token constructions.

Let the input be a sequence of $L$ token vectors collected in the rows of
$X\in\R^{L\times d}$. Rows index token positions, while columns index embedding
coordinates. Attention forms queries, keys, and values through
learned projections,
\[
Q=XW_Q,\qquad K=XW_K,\qquad V=XW_V,
\]
with $W_Q,W_K\in\R^{d\times d_k}$ and $W_V\in\R^{d\times d_v}$. It then returns
\begin{equation}
\mathrm{Attention}(Q,K,V)
=
\mathrm{softmax}\!\left(\frac{QK^\top}{\sqrt{d_k}}\right)V.
\label{eq:attn}
\end{equation}
The softmax is applied to each row of its matrix argument. For a vector
$z=(z_1,\dots,z_L)^\top$, it is defined by
\begin{equation}
\mathrm{softmax}(z)_i
=
\frac{\exp(z_i)}{\sum_{j=1}^{L}\exp(z_j)},
\qquad i=1,\dots,L.
\label{eq:softmax}
\end{equation}
Each row of the resulting weight matrix is therefore nonnegative and sums to
one. Each output row of \eqref{eq:attn} is consequently a convex combination of
the rows of $V$.

Three properties are worth isolating. First, the row weights in \eqref{eq:attn}
are data dependent. Unlike fixed regression coefficients, they are
recomputed for each input from that input's own content. Second, the map is
global. Since rows of $X$ index token positions, any
two rows can interact directly through the $L\times L$ score matrix $QK^\top$.
Dependence between distant token positions therefore need not pass through
intermediate states. {This is a property of the computational graph within the supplied context,
not an exclusive ability to use the remote past. A finite VAR conditions
directly on finitely many lags, while VARMA, state-space, recurrent, and
long-memory specifications can transmit information from earlier dates
\citep{lutkepohl2005new,hamilton1994time}.} Third, the
normalization by $\sqrt{d_k}$ is substantive. When the entries of $Q$ and $K$
are approximately independent with unit variance, the inner products
$q^\top k$ have variance of order $d_k$. Without this scaling, the
softmax can put almost all weight on a single key, making the attention weights
hard to adjust during training.

\begin{figure}[t]
\centering
\includegraphics[width=0.95\textwidth]{ 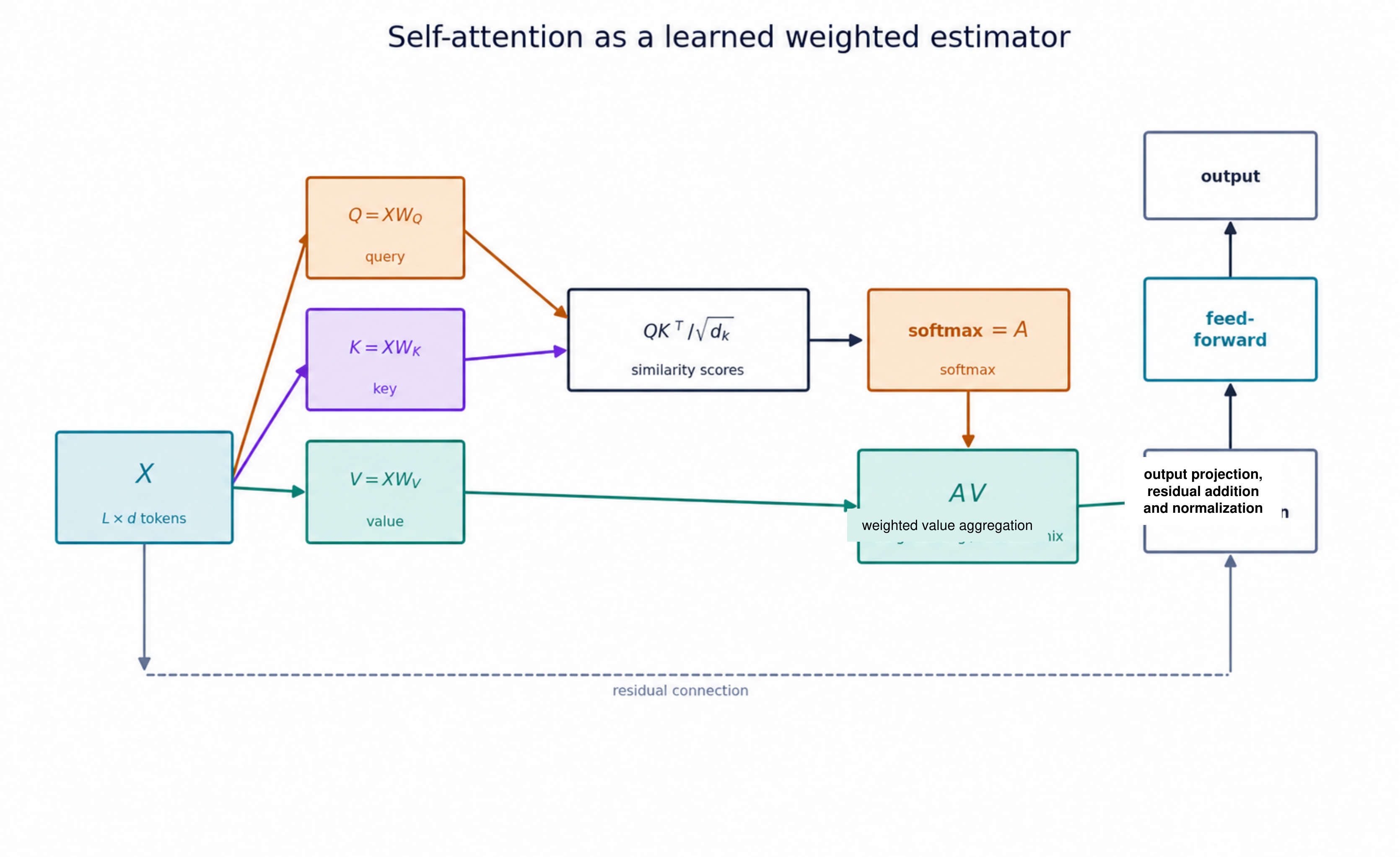}
\caption{{Self-attention as learned weighted aggregation. Queries and
keys determine a row-stochastic weight matrix $A$, and $AV$ combines the
learned value vectors across tokens. The output projection maps the attention output to the same dimension
as the input. A residual connection adds the input to this projected
output. The block also includes layer normalization and a feed-forward
network. The kernel-smoothing analogy concerns normalized
similarity weights. The factor-model analogy is limited: $V$ need not be a
small set of identified common factors, and input-dependent rows of $A$ are
not fixed factor loadings. The schematic does not imply equivalence to a
low-rank factor model.}}
\label{fig:attnarch}
\end{figure}

A complete transformer block applies \eqref{eq:attn} in $M$ parallel heads,
typically with head dimension $d/M$. The head outputs are concatenated and mixed
by an output projection. A position-wise feed-forward network, residual
connections, and layer normalization complete the standard block. Because attention itself is permutation equivariant, ordering is supplied
{separately through positional information. An autoregressive causal mask sets
scores for later positions to $-\infty$, so their softmax weights are zero.
A forecasting encoder whose entire input precedes the forecast origin may
instead attend bidirectionally within that observed window without leaking
future targets.} Figure~\ref{fig:attnarch} summarizes the standard transformer block.

The attention operation also admits two useful econometric readings, as previewed in
Figure~\ref{fig:attnarch}: attention as a learned weighted estimator and
attention as adaptive pooling.
For a single query $q$, the output of \eqref{eq:attn} has the form of a
Nadaraya--Watson kernel smoother over keys $k$. The query plays the role of the evaluation
point, the keys play the role of design points, and the values are the
quantities being averaged. The similarity kernel is
\[
\kappa(q,k)=\exp(q^\top k/\sqrt{d_k}).
\]
Attention differs from a classical smoother because similarity is not measured
in the raw input space. The matrices $W_Q$ and $W_K$ learn the representation in
which tokens are compared. The quantities being
averaged are also learned through $W_V$. 
\par\begingroup\ignorespaces 

The second interpretation is adaptive pooling. Writing \eqref{eq:attn}
in matrix form gives
\[
\mathrm{Attention}=AV,
\qquad
A=\mathrm{softmax}(QK^\top/\sqrt{d_k}), \qquad
V=XW_V .
\]
Here $A$ is a row-stochastic $L\times L$ weight matrix. Each row of $A$ gives
the weights used at one token position, and the rows of $V$ are the learned
value vectors being pooled. The operation therefore combines information across
tokens through weights that depend on the current input. This is related to the
pooling logic of factor models, but the analogy is limited. Dynamic factor
models \citep{stock2002forecasting,baing2002} summarize a high-dimensional
panel through a small number of common components with loadings fixed after
estimation. Attention instead forms input-dependent weighted averages of value
vectors. The resulting representation need not be low-dimensional. 
\par\endgroup

\subsection{From observations to forecasts}
\label{sec:attn-forecast}

\par\begingroup\ignorespaces 

Suppose the $p$ series in $\bm y_s\in\mathbb R^p$ are observed at each
date $s$. To forecast $\bm y_t$, we use the $L$ observations ending at
$t-1$, with one token per date.
{Let $\varphi:\mathbb R^p\to\mathbb R^d$ denote the embedding
map and reserve $\mathbb E$ for expectation.} Each token
$x_s=\varphi(\bm y_s)$ is a $d$-dimensional column vector. Stacking the
tokens as rows gives the input matrix
\[
X_t=
\begin{bmatrix}
x_{t-L}^{\top}\\
\vdots\\
x_{t-1}^{\top}
\end{bmatrix}
\in\mathbb R^{L\times d}.
\]

For a single attention head, the latest token supplies the query, and
each lag $i=1,\ldots,L$ supplies a key and a value, given by
\[
q_t=W_Q^\top x_{t-1},\qquad
k_{t-i}=W_K^\top x_{t-i},\qquad
v(\bm y_{t-i})=W_V^\top x_{t-i}.
\]
The matrices $W_Q,W_K\in\mathbb R^{d\times d_k}$ give queries and keys
with $d_k$ coordinates, so their inner product is a scalar.
The matrix $W_V\in\mathbb R^{d\times d_v}$ gives each value $d_v$
coordinates. Across the full window, $Q_t=X_tW_Q$ and $K_t=X_tW_K$
are $L\times d_k$, and $V_t=X_tW_V$ is $L\times d_v$.

Let $a_i(\bm y_{t-L:t-1})$ be the weight the latest query assigns to
lag $i$. The head averages the values to produce a vector $\bm z_t$.
A learned matrix $D\in\mathbb R^{p\times d_v}$ then maps that vector
to the $p$ series being forecast, giving
\begin{equation}\label{eq:attn-var}
\bm z_t=\sum_{i=1}^{L}a_i(\bm y_{t-L:t-1})\,v(\bm y_{t-i})
\in\mathbb R^{d_v},
\qquad
\widehat{\bm y}_t=D\bm z_t\in\mathbb R^p.
\end{equation}
The value dimension $d_v$ is chosen as part of the architecture, whereas
$p$ is determined by the forecasting target.
The weights are
\begin{equation}
a_i(\bm y_{t-L:t-1})
=
\frac{
\exp\!\left(q_t^\top k_{t-i}/\sqrt{d_k}\right)
}{
\sum_{j=1}^{L}\exp\!\left(q_t^\top k_{t-j}/\sqrt{d_k}\right)
}.
\label{eq:aweight}
\end{equation}

{The scalar score $q_t^\top k_{t-i}/\sqrt{d_k}$ determines the
relative weight assigned to lag $i$. Although the query uses only
the latest token here, each attention weight depends on the
full window through the softmax denominator. In a stacked
transformer, the representation at the latest position has
already combined information from earlier tokens, so the
query can also depend on the wider history.}

{Both the VAR in~\eqref{eq:var} and the attention forecast
in~\eqref{eq:attn-var} sum over lags. In the VAR, each lag has its own
coefficient matrix $A_i$, whose entries may have either sign. In a single
attention head, each $a_i$ is a nonnegative scalar. These weights sum to
one across lags. Each lag therefore contributes $a_i D\,v(\bm y_{t-i})$
to the forecast, with the same value map $v$ and forecasting matrix $D$
at every lag. The weights vary with the observed window, so the forecast
can depend nonlinearly on the past even when $v$ is linear.} {Figure~\ref{fig:var2attn} illustrates how the same fitted
attention model can give different weights to the lags in seasonal and
trending windows.}
\begin{figure}[t]
\centering

\includegraphics[width=0.98\textwidth]{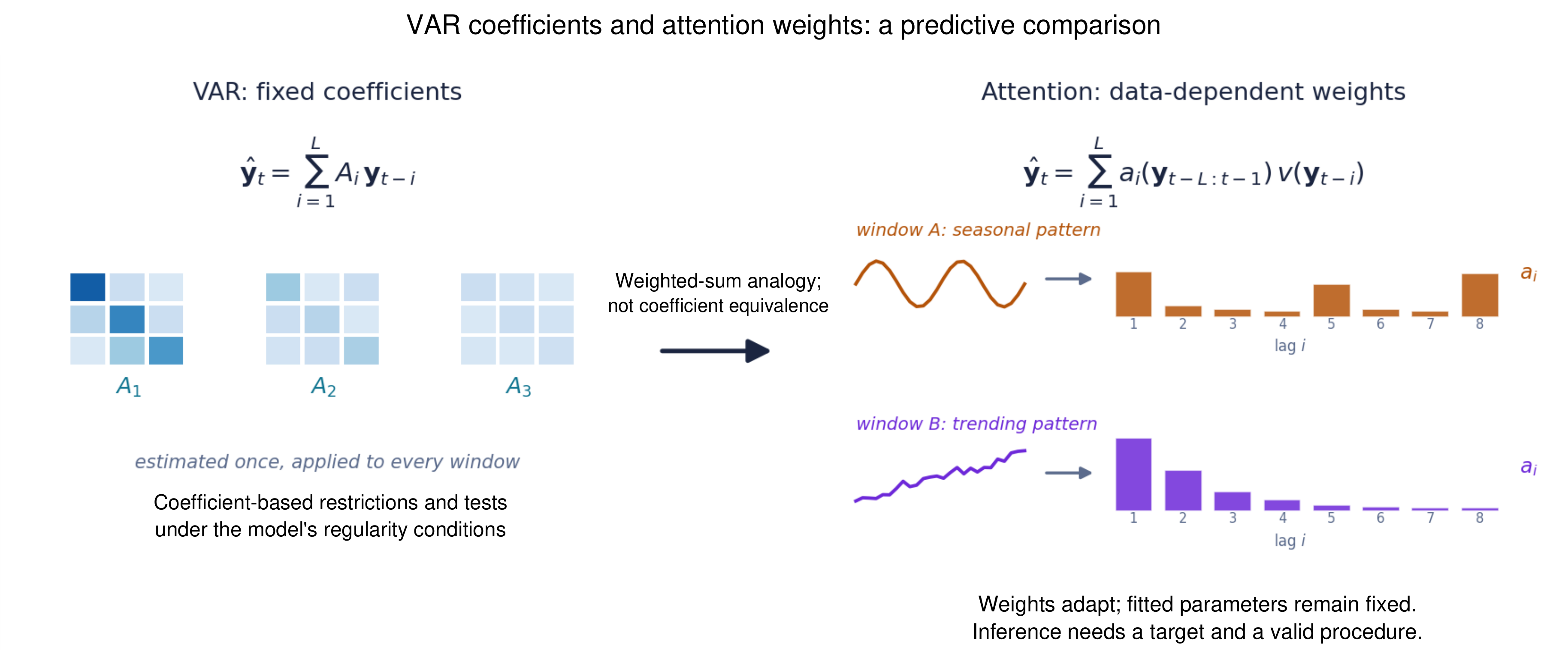}

\caption{{Comparing VAR and single-head attention forecasts.
Left: the VAR applies the same lag matrices $A_i$ to every window.
Right: attention applies the same value transformation $v$ at every
lag. The scalar weights $a_i$ change with the window, and the shared
matrix $D$ maps the weighted average to the forecast. The bars show
these weights for two illustrative windows.} {The profiles
are schematic, not empirical estimates. Structural interpretation and
inference are discussed separately in Section~\ref{sec:classical}.}}
\label{fig:var2attn}
\end{figure}

{Equation~\eqref{eq:attn-var} describes a single head. A full
transformer can add positional information, combine multiple heads
through an output projection, and stack blocks with residual
connections and feed-forward layers.}
{Lag ordering requires information beyond the observed values.
In Equation~\eqref{eq:aweight}, the index $i$ labels the lag, but the key
$k_{t-i}$ is computed from $\varphi(\bm{y}_{t-i})$ alone. Selecting
the latest token as the query identifies the most recent
position, but does not generally distinguish the order of
the remaining tokens. One way to include this information is
to associate each lag $i$ with a position vector
$\bm{r}_i\in\mathbb{R}^{d}$ and define
\[
q_t=W_Q^\top\{\varphi(\bm{y}_{t-1})+\bm{r}_1\},
\qquad
k_{t-i}=W_K^\top\{\varphi(\bm{y}_{t-i})+\bm{r}_i\}.
\]
Thus, identical observations at different lags can receive
different attention scores. The position vectors may be
specified in advance or estimated jointly with the other
parameters. In either case, they remain fixed when forecasting
new windows.}

{
For point forecasting, the embedding and projection parameters
can be estimated jointly by minimizing
\[
\mathcal{L}(\theta)
=\frac{1}{|\mathcal{T}|}
\sum_{t\in\mathcal{T}}
\|\bm{y}_t-\widehat{\bm{y}}_t(\theta)\|_2^2,
\]
where $\mathcal{T}$ contains the training forecast dates and
$\theta$ collects the embedding parameters, the projections $W_Q,W_K,W_V$,
the forecasting matrix $D$, and any learned positional parameters.
After estimation, these parameters remain fixed, while the embeddings,
queries, keys, values, and attention weights are recalculated for each
window.}

{
With $M$ heads, the same input window is processed through
$M$ sets of projection matrices. Head $m$ has its own
$W_Q^{(m)},W_K^{(m)}\in\mathbb{R}^{d\times d_k}$ and
$W_V^{(m)}\in\mathbb{R}^{d\times d_v}$, and therefore its own
attention weights and value transformation. Keeping the value
dimension $d_v$ for each head, its output is
\[
\bm{z}_t^{(m)}
=\sum_{i=1}^{L}
a_i^{(m)}(\bm{y}_{t-L:t-1})\,
v^{(m)}(\bm{y}_{t-i}),
\qquad
v^{(m)}(\bm{y}_{t-i})
=(W_V^{(m)})^\top \varphi(\bm{y}_{t-i})
\in\mathbb{R}^{d_v}.
\]
Each head can assign different weights to the lags and use
different combinations of the embedded variables. Together,
the heads provide several learned representations of the same
observed history. For a direct $H$-step forecast, their outputs
are combined as
\[
\widehat{\bm{y}}_{t+h-1}
=\sum_{m=1}^{M}D_{h,m}\bm{z}_t^{(m)},
\qquad
D_{h,m}\in\mathbb{R}^{p\times d_v},
\qquad h=1,\ldots,H.
\]
Each $D_{h,m}$ maps the $d_v$ coordinates of head $m$ to the $p$
variables forecast at horizon $h$. With one head and one horizon,
$D_{1,1}=D$ recovers the one-step forecast in~\eqref{eq:attn-var}.
Different horizons can therefore use different combinations of the head
outputs, while all forecasts use the same observed window ending at $t-1$.
Equivalently, concatenating the $M$ head outputs gives a vector in
$\mathbb R^{M d_v}$. The block matrix $[D_{h,1}\ \cdots\ D_{h,M}]$
has dimension $p\times (M d_v)$ and maps this vector to the same forecast.

The embedding parameters, all head-specific projection matrices, any learned
positional parameters, and the matrices $D_{h,m}$ are estimated jointly.
Let $\theta$ collect these parameters. For point forecasting, one possible
training criterion is
\[
\mathcal{L}_H(\theta)
=
\frac{1}{H|\mathcal{T}_H|}
\sum_{t\in\mathcal{T}_H}
\sum_{h=1}^{H}
\left\|
\bm{y}_{t+h-1}
-\widehat{\bm{y}}_{t+h-1}(\theta)
\right\|_2^2,
\]
where $\mathcal{T}_H$ contains forecast dates whose input windows
and all $H$ outcomes lie within the training sample.}
In stacked blocks, the parameter vector also includes the
feed-forward and normalization parameters. Hyperparameters are selected on
earlier validation dates. Quantile or distributional forecasting requires
an appropriate output specification and loss. Attention alone does not
supply them.

The training loss may also include explicit penalties, including LASSO-type
terms, as discussed in Section~\ref{sec:transformer-adaptations}.
The forecasting origin remains $t-1$ for all horizons in this subsection.
In the running macro example, the attention weights in \eqref{eq:attn-var}
may give different importance to past inflation or interest-rate episodes
depending on the current state of the economy. A recent high-inflation
window and a tranquil window can therefore induce different effective lag
profiles from the same fitted model.
{Attention weights describe the model's predictive
aggregation. They are not themselves impulse responses or
Granger-causality measures, whose interpretation and inference
require the specifications discussed in Section~\ref{sec:classical}.}

\subsection{Tokenizing time series}
\label{sec:tokenize}

Time-series transformers inherit the token representation from natural
language processing. In language modelling, a token is the elementary
unit consumed by the model. Since neural networks compute on vectors rather than
symbols, raw text must first be segmented into discrete units. Modern systems
typically use subword units, so a frequent word may be one token while a rare
word is assembled from several pieces.
In this language pipeline, two steps should be kept distinct. Tokenization
assigns a discrete index, which
carries no meaning by itself. Embedding maps that index to a learned vector, and
the usable representation resides in this vector space. A sentence therefore
becomes a sequence of embeddings, stacked as $X\in\R^{L\times d}$. This array is
the input to the attention operation in \eqref{eq:attn}
(Figure~\ref{fig:nlptoken}).

\begin{figure}[t]
\centering
\includegraphics[width=0.95\textwidth]{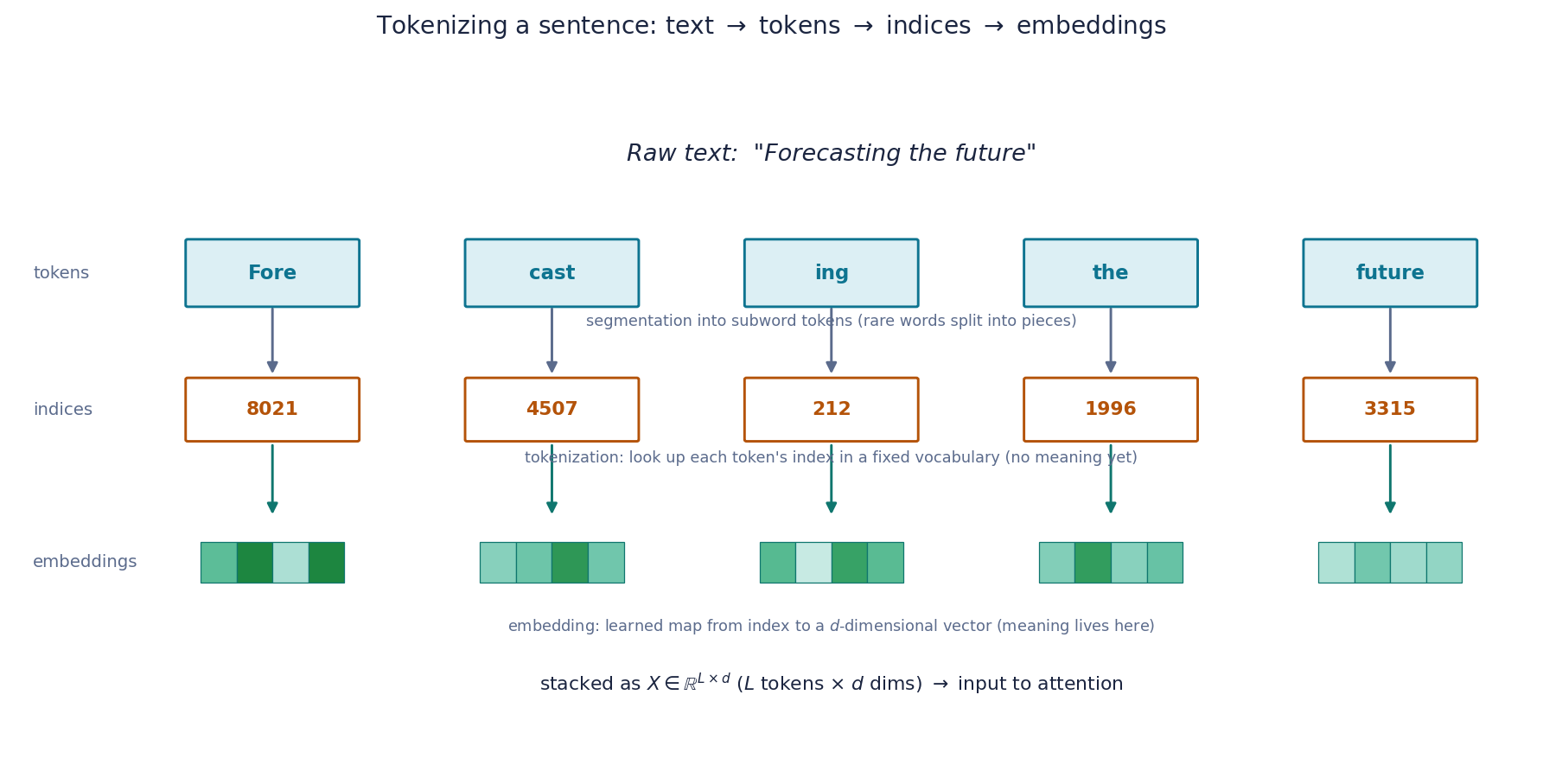}
\caption{Tokenization in language modelling. In this illustration, the word
``Forecasting'' is split into the subword tokens \texttt{Fore}, \texttt{cast},
and \texttt{ing}. Each token receives a vocabulary index, which the embedding
layer maps to a learned vector. Stacking these vectors as rows gives the
input to attention, $X\in\R^{L\times d}$. Figure~\ref{fig:tokenize} shows
how time-series foundation models can use quantized values or real-valued
patches as tokens.}
\label{fig:nlptoken}
\end{figure}

For a real-valued series there is no canonical token. The earliest
time-series transformers treated a single observation as a token.
With time-point tokens, \eqref{eq:attn} constructs an $L\times L$ matrix
of scores, so its computational cost scales quadratically with the
context length. This becomes costly in forecasting applications, where
long look-back windows are often used to capture seasonality, cycles,
and other persistent dependence. An early wave of time-series
transformers therefore focused on reducing this cost.
\citet{zhou2021informer} retain only the most informative queries,
reducing the cost to order $L\log L$, and decode the forecast horizon
in a single pass.
\citet{wu2021autoformer} embed a seasonal--trend decomposition and replace
inner-product attention with an auto-correlation operator.
\citet{zhou2022fedformer} move the computation to a frequency basis and
obtain linear complexity. These models reported improvements over the
unmodified transformer on standard long-horizon benchmarks, but the
gains proved less settled than they first appeared.

The limits of these early transformer forecasters were made explicit by
\citet{zeng2023transformers}.
They showed that DLinear, which applies separate linear maps to
the trend and seasonal components and sums their forecasts,
matched or outperformed these architectures on the evaluated
benchmarks. Their analysis questioned how effectively the
tested transformers exploit temporal order and longer
look-back windows.
These comparisons show that more elaborate architectures do not
necessarily outperform simple linear baselines. They motivate closer
attention to the input representation, alongside decoding and
evaluation design, but do not isolate the effect of tokenization.
The following approaches explore alternative ways of constructing
time-series tokens.

\citet{nie2023patchtst} propose PatchTST, which groups $P$ contiguous
observations of each variable into a single token.
{The resulting
$P$-dimensional vector is mapped to a $d$-dimensional embedding. Each token
then carries local shape rather than a single value, and attention compares
different time segments of that variable. For nonoverlapping patches with
$P$ dividing $L$, the token count is $L/P$. With stride $S$ and $L\geq P$,
the unpadded token count is $N=1+\lfloor(L-P)/S\rfloor$.
Patching reduces the computational cost of attention, but the model
must still compute patch embeddings and perform its other operations.}
{PatchTST also adopts channel
independence, applying a shared univariate model to each series, including
shared embedding and attention parameters. This avoids modelling
cross-series dependence directly and may reduce overfitting
in some settings, but it can also discard useful predictive
information from other series. PatchTST shares parameters across series
but forecasts each series using only its own history, rather than
directly conditioning on the histories of other variables.}

{Multivariate patches provide another construction. With patches
of length $P$ across all $p$ series, the embedding maps
$\mathbb R^{pP}$ to $\mathbb R^d$. With the same stride $S$, the token count
is given by the same expression for $N$.}
\citet{liu2024itransformer} propose iTransformer, in which each
variable's history within the input window is treated as a token, so
attention runs across the cross-section rather than across time.
Temporal structure is then handled mainly inside the feed-forward component.
{Each variable's history is first mapped to a $d$-dimensional
token vector. Stacking these $p$ token vectors as rows gives an input
array in $\mathbb R^{p\times d}$. Self-attention on these tokens produces
a $p\times p$ weight matrix. Each variable's resulting representation
is used for forecasting.}
{PatchTST processes channels separately while sharing parameters,
whereas iTransformer makes cross-variable interactions the object of
attention. Neither design is uniformly preferable.}

Another approach follows the language pipeline more literally.
\citet{ansari2024chronos} first scale the observations and then assign
each scaled value to one of a fixed set of numerical intervals (bins).
Each bin is represented by a discrete token, so the series becomes a
sequence of tokens that can be used to train a standard language-model
architecture. Binning limits numerical precision because values in the
same bin receive the same token.

Quantized vocabularies and real-valued patches are the two constructions
that organize much of the foundation-model literature, and they are
contrasted in Figure~\ref{fig:tokenize}.


The choice is not merely computational. The token fixes the objects over which
attention is computed. For the macro panel, the model may compare individual
monthly observations, short patches of local dynamics, or entire variable
histories. A patch token may encode a local inflation cycle. A variable token may
ask how the inflation series as a whole relates to the policy-rate series. The
token therefore fixes the economic object being compared.

\begin{figure}[t]
\centering
\includegraphics[width=0.95\textwidth]{ 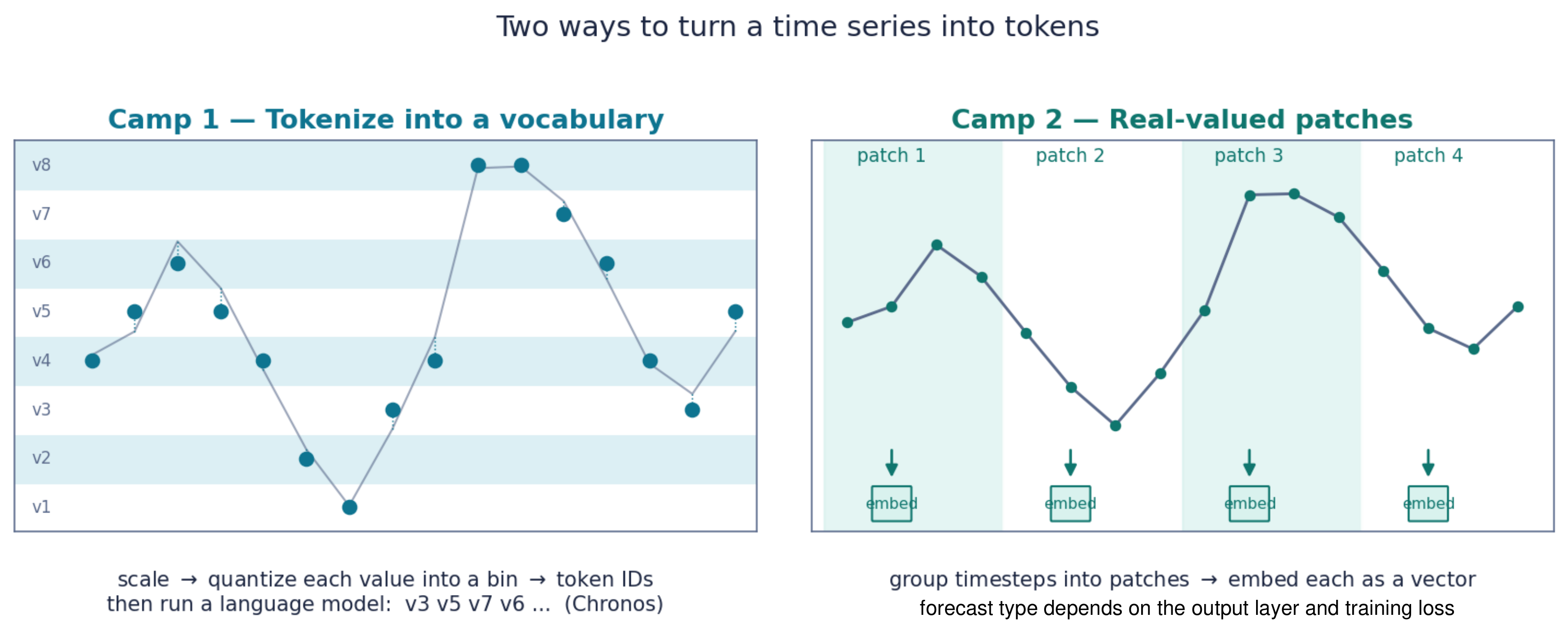}
\caption{{Two time-series representations. Left: quantization maps
scaled observations to vocabulary bins. Right: real-valued patches group
consecutive observations before embedding. Quantization introduces a finite
resolution. Patching changes the representation and token count. Neither
representation alone determines whether forecasts are probabilistic: that
depends on the forecasting layer, predictive distribution, and training objective.}}
\label{fig:tokenize}
\end{figure}

\subsection{Architectural adaptations for time-series forecasting}
\label{sec:transformer-adaptations}
Beyond tokenization, time-series transformers have also been adapted
to address distribution shift.
\citet{kim2022revin} propose reversible instance normalization (RevIN),
which removes the location and scale of each window before modelling and
restores them afterwards. It is a neural counterpart of the detrending,
demeaning, and rescaling that classical practice applies before estimation.
Later work models the shift rather than only removing it, feeding information
from the raw series and its statistics back into the attention computation
\citep{liu2022nonstationary}.

\par\begingroup\ignorespaces 
Attention constructs input-dependent interactions using shared
learned parameters. With variable tokens, for example, the
pairwise attention weights are generated by common projection
matrices rather than estimated as unrelated coefficients
\citep{liu2024itransformer}. This parameter sharing imposes
structure on the forecasting function, but does not remove
the need to control model complexity. Transformer architectures
can therefore be combined with explicit regularization,
including LASSO-type penalties on selected trainable network
weights. Architecture determines how information is combined.
Regularization constrains how the resulting model is estimated.
With temporal tokens, attention also permits direct interactions
between distant positions and parallel computation across the
observed context \citep{vaswani2017attention}.
\par\endgroup 

\par\begingroup\ignorespaces 

\citet{antolin2026attention} develop attention-based vector
autoregressions in which attention governs the conditional mean
and variance, with dynamics determined by past observations.
Their framework nests constant-parameter VARs and approximates
dynamic factor models, and is accompanied by Bayesian priors
and estimation procedures designed for macroeconomic and
financial samples. This provides a concrete example of
combining attention with econometric structure, rather than
treating the two as competing approaches. The reported nesting
and approximation properties concern their specified model.
They should not be attributed to generic transformer
architectures.
\par\endgroup 

\par\begingroup\ignorespaces 

Read against the three challenges, the transformer era is uneven.
High dimensionality is addressed partly through architectural
choices. Channel independence avoids modelling cross-series
dependence directly while sharing parameters across series,
whereas cross-variable attention constructs input-dependent
interactions between series
\citep{nie2023patchtst,liu2024itransformer}.
Tokenization also
shapes the problem: point-wise, patch-wise, and variable-wise
tokens define different objects over which attention is
computed. These architectural choices can be complemented
by explicit regularization.
\par\endgroup 

\par\begingroup\ignorespaces 

For nonstationarity, unit-root behaviour, cointegrating
relations, structural breaks, and parameter drift require
distinct modelling assumptions. Regime dependence need not
itself imply nonstationarity. 
Normalization may mitigate
shifts in location and scale, but does not by itself impose
a unit-root structure, cointegrating restrictions, or an
error-correction mechanism.
The change-point and locally stationary frameworks discussed
in Section~\ref{sec:classical} provide complementary approaches
to modelling instability.
\par\endgroup 

\par\begingroup\ignorespaces 
Nonlinearity is where the architectural contribution is most
direct. Equation~\eqref{eq:attn} combines observations through
input-dependent weights, while the feed-forward layers provide
nonlinear transformations \citep{vaswani2017attention}.
Together, these components represent nonlinear predictive
relationships through a shared architecture. Forecasting gains
nevertheless depend on the specification and available training
information.
\par\endgroup 

The token representations in Section~\ref{sec:tokenize} also provide a natural
bridge from time-series transformers to foundation models.
Section~\ref{sec:fm} examines how models trained across collections of series
use such representations for forecasting new series.

\section{Foundation models and zero-shot forecasting}
\label{sec:fm}
Until recently, forecasting was usually organized around a per-data-set
estimation workflow. For each application, the researcher selected a model,
estimated its parameters on the available series, and then re-estimated the
model as new data arrived. This workflow is difficult when the target series is
short, because there is little data from which to estimate the model. Foundation
models change the order of operations. The model is first pretrained on a large
and heterogeneous collection of series. In the zero-shot setting considered
here, it is then applied to a new series by
conditioning on that series' observed history, without re-estimating the model's
parameters. In this sense, forecasting becomes less an exercise in estimating a
new model for each data set and more an exercise in using a pretrained model
with the right context.

The analogy with language modelling is direct. A language model is pretrained on
many texts and then uses the prompt to generate a continuation. A time-series
foundation model is pretrained on many series and then uses the observed history
to generate a forecast. The time-series case, however, is harder in several
ways. Series differ in scale, sampling frequency, seasonal pattern, and degree
of nonstationarity. A model trained across many domains must therefore learn how
to compare and use histories that may live on very different numerical scales.

In the macro example, zero-shot forecasting would mean applying a pretrained
model to a new inflation-output-interest-rate panel without re-estimating its
parameters.

Zero-shot forecasting refers here to prediction without target-specific
parameter estimation. The observed history is supplied as input to the
pretrained model, which produces a forecast without updating its weights.
Whether the target series or dataset was excluded from pretraining
is a separate evaluation consideration.

In-context learning describes task adaptation through supplied context
while the model's parameters remain fixed. Conditioning on history alone,
however, does not establish a particular learning mechanism.
Recent theory outside time series interprets in-context learning as
implicit Bayesian inference. Under the modelling assumptions of these
analyses, 
the prompt updates beliefs under a prior represented in
the pretrained model, and risk can fall as context length grows
\citep{xie2022icl,muller2022pfn,wies2023learnability}. These results,
however, rely on specific assumptions about task mixtures and dependence
between examples. A forecasting context is a single dependent path,
often nonstationary. Applying these arguments to that setting requires
additional justification.

Context length is important for both zero-shot forecasting and
in-context learning. A longer context lets the model condition on more
of the target series. In multivariate settings, it can also include
related series supplied alongside the target. The forecast then depends
not only on what the model learned during pretraining, but also on how
much relevant history is available at prediction time.

The pretrained time-series foundation models in Table~\ref{tab:fm}
can be grouped according to the tokenization methods described in
Section~\ref{sec:tokenize}. These models are pretrained across collections
of series for reuse on new series, and many use a Transformer backbone.
Chronos \citep{ansari2024chronos} represents the quantized-vocabulary
approach. It uses a standard language-model architecture and treats
forecasting as classification over value bins. This makes the connection
to language modelling especially direct, but binning limits numerical
precision.
TimesFM \citep{das2024timesfm} and Moirai \citep{woo2024moirai} represent
the real-valued-patch approach. TimesFM is a decoder-only, long-context
model. Moirai is an any-variate, mixed-frequency model designed to handle
different numbers of variables and different sampling frequencies.
Real-valued patches have become a common design because they are compact,
fast, and naturally suited to continuous-valued prediction.
Other designs modify this template. Lag-Llama \citep{rasul2023lagllama}
encodes each date through lagged values. TimeGPT \citep{garza2023timegpt}
is an early closed-weight commercial system.
Time-MoE \citep{shi2024timemoe} pursues mixture-of-experts scaling.

\begin{table}[t]
\centering
\caption{Time-series foundation models discussed in the text, classified by tokenization,
architecture, predictive output, and availability of weights. The two predominant tokenizations,
quantized vocabularies and real-valued patches, are developed in Section~\ref{sec:tokenize}.}
\label{tab:fm}
\small
\begin{tabular}{@{}lllll@{}}
\toprule
Model & Tokenization & Architecture & Output & Weights \\
\midrule
Chronos    & quantized vocabulary  & encoder--decoder      & categorical    & open \\
TimesFM    & real-valued patches   & decoder-only          & quantiles      & open \\
Moirai     & real-valued patches   & any-variate encoder   & distributional & open \\
Lag-Llama  & lagged covariates     & decoder-only          & distributional & open \\
TimeGPT    & undisclosed           & encoder--decoder      & quantiles      & closed \\
Time-MoE   & point-wise real values & decoder-only MoE & point & open \\
\bottomrule
\end{tabular}
\end{table}

\par\begingroup\ignorespaces 

In the GIFT-Eval snapshot
\citep{aksu2024gifteval}, each of the three zero-shot foundation
models in Table~\ref{tab:gifteval} has a lower aggregate normalized
weighted quantile loss than every statistical or neural baseline
listed. The comparison uses the same 97 tasks for all models.
The evaluated checkpoints are Moirai-2.0, TimesFM-2.5, and Chronos-2.
Their architectures and forecasting layers may differ from the representative
earlier designs in Table~\ref{tab:fm}.
\par\endgroup 

\par\begingroup\ignorespaces 

Pretraining on related series or on an earlier history of the target is not,
by itself, test-data leakage. The concern is that the model may have already seen the actual outcomes
it is asked to predict or may use other information that was unavailable
when the forecast was made. Evaluation on outcomes
that became available only after the checkpoint's documented training cutoff
helps guard against direct target leakage.

GIFT-Eval's contamination-aware design does not
establish that every evaluated checkpoint is uncontaminated.
For the pretrained models in Table~\ref{tab:gifteval}, the
archived submission metadata reports no test-data leakage.
Appendix~\ref{app:gifteval} documents the available evidence
on training--evaluation separation for each checkpoint and
identifies information that is not publicly recorded.
These statements reflect the available documentation rather
than an independent audit of the complete pretraining data.
\par\endgroup 

\par\begingroup\ignorespaces 

For financial applications, the usefulness of pretraining should be assessed
for the particular target, horizon, and evaluation period, rather than presumed
from general-purpose benchmarks. Evidence that pretraining helps or does not help with a particular
financial forecasting task should not be generalized to finance as a whole. Such evaluations should use held-out test outcomes, document possible
overlap between training and evaluation data, and compare the pretrained
model with competitive econometric models and machine-learning models
trained for the target task. Where feasible, evaluations should compare the same architecture with
and without external pretraining. The application in Appendix~\ref{sec:application} does not include this comparison and
therefore cannot assess the contribution of external pretraining.
\par\endgroup 

\begin{table}[t]
\centering
\caption{Representative GIFT-Eval results \citep{aksu2024gifteval}, computed
from public result files with a reported access date of 14 July 2026. {Both blocks report the same score. On each
of the 97 tasks, the model's weighted quantile loss is divided by the
seasonal-naive loss. The table reports the geometric mean of these
ratios, so seasonal naive scores one and lower values are better.
The upper block lists statistical and neural baselines, and the lower
block lists zero-shot foundation models.} TimesFM-2.5 denotes a later checkpoint
of the model introduced by \citet{das2024timesfm}. The
leaderboard is updated over time, so these values should be read as a dated snapshot.
{Details of the repository version, result files, model
checkpoints, probabilistic outputs, and treatment of missing
results are provided in Appendix~\ref{app:gifteval}.}}
\label{tab:gifteval}
\small
\begin{tabular}{@{}lc@{}}
\toprule
{Model} & {Normalized WQL ($\downarrow$)} \\
\midrule
\multicolumn{2}{@{}l}{{\emph{Statistical and neural baselines}}}\\
\midrule
Seasonal naive                        & 1.000 \\
Auto-ARIMA \citep{hyndman2008automatic} & 0.912 \\
DeepAR \citep{salinas2020deepar}      & 0.853 \\
DLinear \citep{zeng2023transformers}  & 0.846 \\
N-BEATS \citep{oreshkin2020nbeats}    & 0.816 \\
\midrule
\multicolumn{2}{@{}l}{\emph{Zero-shot foundation models}}\\
\midrule
Moirai-2.0 \citep{liu2025moirai2}     & 0.516 \\
TimesFM-2.5 \citep{das2024timesfm}    & 0.490 \\
Chronos-2 \citep{ansari2024chronos2}  & 0.485 \\
\bottomrule
\end{tabular}
\end{table}

Against the three challenges, pretraining can ease data limitations in
high-dimensional forecasting. Broad cross-corpus training lets the model
pool information across many domains. Any-variate architectures
\citep{woo2024moirai} allow the number of input series to vary.
{These architectures also impose restrictions and can be
combined with explicit regularization. Pretraining changes the
information used for estimation rather than eliminating the need
to control model complexity.}
For nonstationarity, scaling and normalization help models operate
across heterogeneous series, but they do not replace the econometric
apparatus of unit roots, cointegration, and error correction.
{Nonlinearity depends on the backbone and forecasting layer.}
{Cross-corpus pretraining can complement application-specific
estimation or support forecasting with fixed parameters. Its value
under dependence and distribution shift must be assessed for the
target task.}
Recent work combines pretrained backbones with generative output
heads based on flow matching or diffusion.
Section~\ref{sec:diffusion} turns to the generative forecasting
mechanism, focusing on how diffusion models sample joint future
paths conditional on the observed history.

\section{Diffusion and generative forecasting}\label{sec:diffusion}
A useful forecast describes a range of possible futures, not just a single path. Decisions about inventory, capacity, and risk rely on quantiles and scenarios rather than a single point. Foundation models that output separate quantiles describe marginal uncertainty
at each horizon, leaving dependence across horizons and series unspecified.
Generative models represent a joint predictive distribution and sample whole future paths (Figure~\ref{fig:why-gen}). Among generative methods, diffusion has emerged as a leading framework for time-series forecasting, building on earlier successes in image and audio synthesis.

In the running example, a diffusion forecaster would not merely produce separate
quantiles for future inflation, output growth, and the policy rate. It would sample
joint future paths, so that a high-inflation scenario comes with a corresponding
path for output and interest rates. The attraction is the explicit representation of dependence across variables
and horizons. The econometric question is whether those scenarios respect the structural and long-run
relations thought to govern the macroeconomic system.
\begin{figure}[t]\centering
  \includegraphics[width=0.95\textwidth]{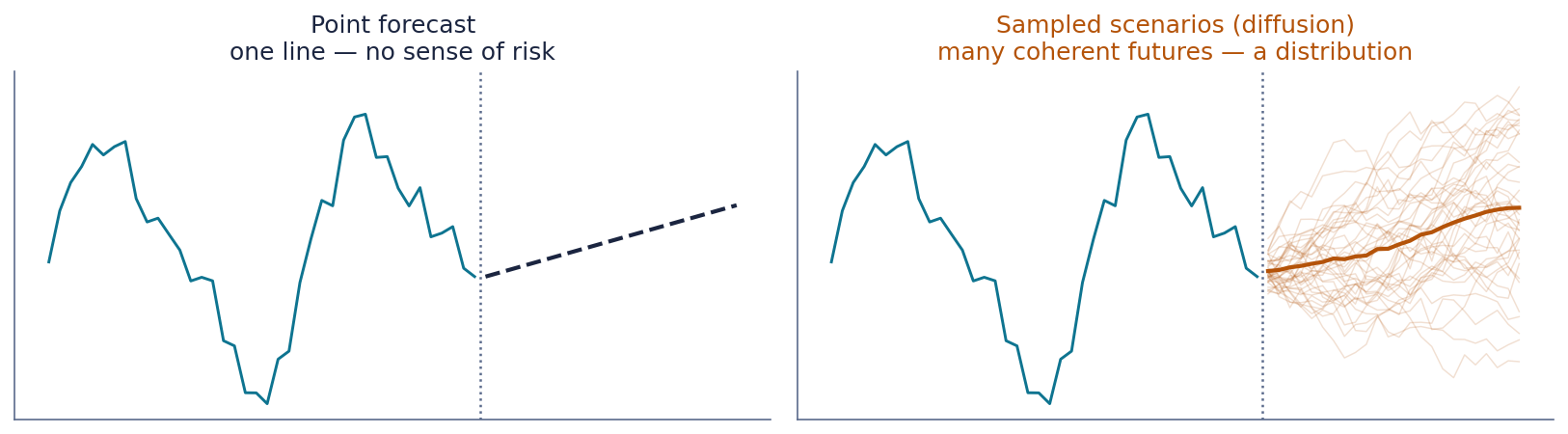}
  \caption{Why generative forecasting. A point forecast is a single trajectory and conveys little information about uncertainty (left). {A conditional generative model samples future paths from its fitted predictive distribution (right). Sampling alone does not establish calibration, realistic dependence, or structural coherence.}}
  \label{fig:why-gen}
\end{figure}

At forecast date $t$, both the VAR and the diffusion model target the
conditional distribution of the same future path given the observed
history $\bm y_{1:t}$. Write this path as
\[
\bm y_{t+1:t+H}
=
(\bm y_{t+1}^{\top},\ldots,\bm y_{t+H}^{\top})^{\top}
\in\mathbb R^{pH}.
\]

The VAR in \eqref{eq:var} can generate a joint draw of this path
by drawing a $p$-dimensional innovation at each future date and
using the simulated observations as inputs at subsequent dates.
Equation~\eqref{eq:var} allows non-Gaussian innovations, which can
produce a non-Gaussian predictive distribution even with linear
dynamics.

For an explicit benchmark, consider a VAR or VARMA with Gaussian
innovations and fixed parameters. If the initial state is known
or has a Gaussian filtering distribution, the joint predictive
distribution is
\begin{equation}\label{eq:varma-pred}
\bm y_{t+1:t+H}\mid\bm y_{1:t}
\sim
\mathcal N_{pH}\!\left(\bm\mu_{t,H},\Sigma_{t,H}\right),
\end{equation}
where $\bm\mu_{t,H}\in\mathbb R^{pH}$ is the conditional mean and
$\Sigma_{t,H}\in\mathbb R^{pH\times pH}$ contains the conditional
covariances across variables and forecast horizons. 

A diffusion model represents the conditional distribution of the
same future path as
\begin{equation}\label{eq:diff-pred}
\bm y_{t+1:t+H}\mid\bm y_{1:t}
\sim
p_\theta\!\left(\,\cdot\mid\bm y_{1:t}\right).
\end{equation}
Here the model learns a denoising procedure for sampling future paths
conditional on the observed history. Compared with the Gaussian
benchmark in \eqref{eq:varma-pred}, this approach can represent a
wider range of predictive distributions, including asymmetric or
multimodal distributions whose shape varies with the observed history.
How well it captures these features depends on the model architecture,
training objective, and sampling procedure.
Density forecasting has been approached with several classes of generative models, including variational autoencoders, normalizing flows, and generative adversarial networks. Each learns to represent the predictive distribution by a different mechanism. We focus here on denoising diffusion, which is the subject of the remainder of this section. Broader reviews of generative forecasting methods are available elsewhere \citep{meijer2024rise,yang2024survey}.


\subsection{Denoising diffusion: mechanism and foundations}
\label{sec:diff-mech}
To fix ideas, we begin with the vanilla diffusion mechanism and introduce forecasting-specific conditioning only in Section~\ref{sec:diff-models}.
Denoising diffusion combines a fixed forward process that gradually corrupts a target sample with a learned reverse process that reconstructs it (Figure~\ref{fig:diff-mech}). The basic idea goes back to \citet{sohldickstein2015}, who formulated generation as the finite-time reversal of a Markov chain that diffuses data into a tractable prior. It became practically useful with the denoising diffusion probabilistic models of \citet{ho2020denoising}. Throughout this subsection, let $x^{(0)}$ denote the target sample to be generated, with the superscript indexing diffusion step rather than calendar time.
In the forward process, each diffusion step shrinks the current state and adds Gaussian noise according to a prescribed variance schedule:
\begin{equation*}
x^{(n)}=\sqrt{\alpha_n}\,x^{(n-1)}+\sqrt{\beta_n}\,\epsilon_n,
\qquad \epsilon_n\sim\mathcal N(0,I).
\end{equation*}
Here $\{\beta_n\}_{n=1}^N$ is a prescribed sequence that controls the amount of noise added at each diffusion step, and $\alpha_n=1-\beta_n$. Writing $\bar{\alpha}_n=\prod_{s=1}^n \alpha_s$, iterating this recursion yields the closed form
\begin{equation}\label{eq:forward}
x^{(n)} = \sqrt{\bar\alpha_n}\,x^{(0)} + \sqrt{1-\bar\alpha_n}\,\epsilon,
\qquad \epsilon\sim\mathcal N(0,I).
\end{equation}
This representation allows training at any diffusion step without simulating the full chain. As $\bar\alpha_n$ approaches zero, the distribution approaches the standard normal prior from which generation begins.

\begin{figure}[t]\centering
  \includegraphics[width=0.95\textwidth]{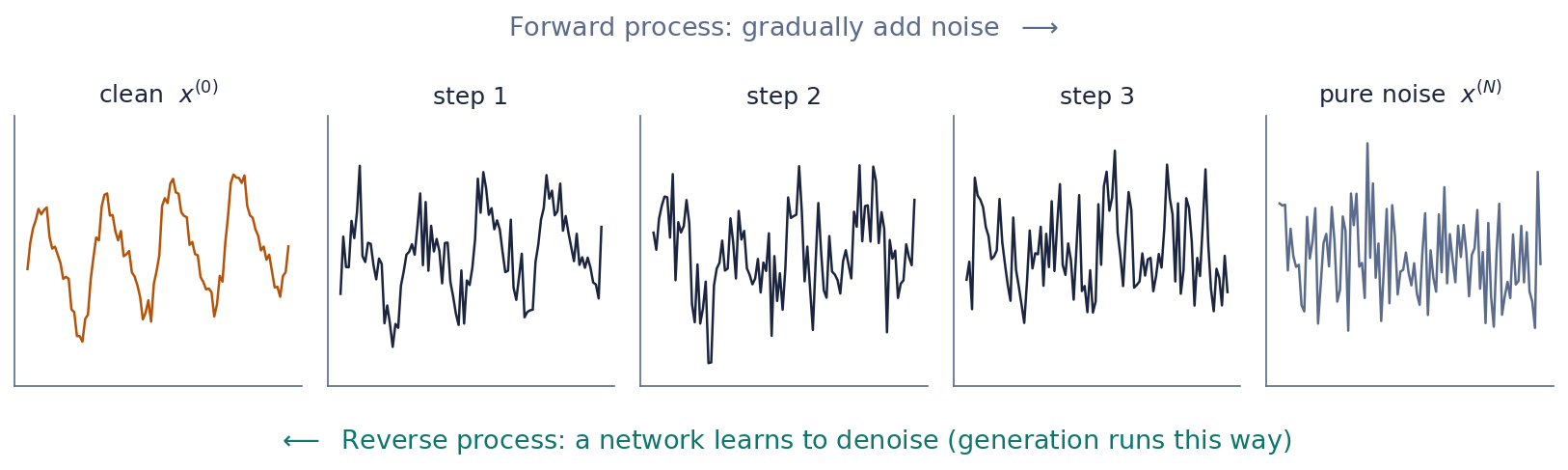}
  \caption{The diffusion mechanism. A fixed forward process gradually adds noise to the series from left to right. A learned reverse process removes that noise from right to left. Generation starts from pure noise and ends in a sample. Conditioning on the observed history turns this generative process into forecasting.}\label{fig:diff-mech}
\end{figure}

The forward process also gives formal content to the analogy drawn in
Section~\ref{sec:classical}, though the comparison should be interpreted with
care. Equation~\eqref{eq:forward} expresses the noised state as an attenuated
initial sample plus a weighted accumulation of Gaussian terms
$\{\epsilon_s\}_{s\le n}$. This shares the moving-average algebra of the Wold
representation, in which the VARMA outcome accumulates past innovations.
The resemblance is formal rather than substantive. VARMA innovations are
disturbances of the data generating process, indexed by calendar time.
Their structural economic interpretation requires identifying assumptions.
Diffusion terms, by contrast, are artificial noise injected by the modeller.
They are indexed by noise level rather than calendar time and are used only
to define the generative mechanism.

What the two constructions share is a linear recursion and the tractability
it affords. For a Gaussian VARMA with fixed parameters and a known or
Gaussian-filtered initial state, that recursion yields a closed-form
predictive density. In diffusion, it yields the closed form
\eqref{eq:forward}, which makes training feasible at arbitrary noise levels.
The key departure is that diffusion learns to invert this accumulation
through a reverse map that can be nonlinear.

For sufficiently small per-step variance, the reverse transition can be approximated by a Gaussian, with a mean predicted by the network. After the noise-prediction reparameterization of \citet{ho2020denoising}, the variational bound yields a timestep-weighted squared-error objective. In practice, one often optimizes the simplified objective
\begin{equation}\label{eq:diff-loss}
  \mathbb{E}_{n,\,x^{(0)},\,\epsilon}
  \big\|\,\epsilon-\epsilon_\theta\!\big(x^{(n)},n\big)\big\|^2,
\end{equation}
where the expectation is taken over a diffusion step $n$ drawn uniformly from $\{1,\dots,N\}$, a clean sample $x^{(0)}$, and Gaussian noise $\epsilon\sim\mathcal N(0,I)$. This objective asks the model to recover the noise injected in the forward process. Under this parameterization, the reverse update draws $x^{(n-1)}$ from a Gaussian with mean
\begin{equation}\label{eq:reverse}
  \mu_\theta\!\big(x^{(n)},n\big)
  = \frac{1}{\sqrt{\alpha_n}}\!\left(x^{(n)}
  - \frac{\beta_n}{\sqrt{1-\bar\alpha_n}}\,\epsilon_\theta\!\big(x^{(n)},n\big)\right),
\end{equation}
so that generation proceeds by repeatedly applying the learned denoiser from level $N$ back to the data.

Up to a known scale factor, predicting the noise is equivalent to estimating the score, that is, the gradient of the log-density of the noised data. This is the perspective developed by \citet{song2019score}, who learn that gradient directly at a range of noise levels and generate samples by annealed Langevin dynamics. In that formulation, generation repeatedly follows the learned score field back toward high-density regions of the data distribution.

These two views, the variational one of \citet{ho2020denoising} and the score-based one of \citet{song2019score}, were unified by \citet{song2021scorebased} within a single stochastic differential equation (SDE) framework. In this formulation, the forward SDE gradually injects noise, while the reverse-time SDE, driven by the learned score, defines the generative process. The same framework also admits a deterministic probability-flow ordinary differential equation (ODE) with the same marginals. In continuous time, this ODE plays the same role as the deterministic denoising diffusion implicit model (DDIM) sampler of \citet{song2021ddim}.

A main practical drawback of classical diffusion sampling is that it requires tens to hundreds of denoising steps. The probability-flow ODE and DDIM already reduce this cost by replacing stochastic reverse sampling with more efficient deterministic trajectories. Subsequent developments push this idea further, including flow matching \citep{lipman2023flow}, which replaces the noising chain with a learned velocity field along a prescribed probability path, and, more recently, consistency models \citep{song2023consistency} and rectified flow \citep{liu2023rectified}. These methods can reduce generation to a handful of function evaluations, making deployment far cheaper.

\subsection{Conditional diffusion for time-series forecasting}
\label{sec:diff-models}

As a reference construction for whole-segment forecasting, let the target
sample be the future segment
\[
x^{(0)} := \bm y_{t+1:t+H}.
\]
In this construction, the forward noising process is unchanged. It perturbs
only this future segment, while the observed history $\bm y_{1:t}$ enters
separately as conditioning information.

In the unconditional setting of Section~\ref{sec:diff-mech}, the reverse
chain learns transitions of the form
$p_\theta(x^{(n-1)} \mid x^{(n)})$.
For forecasting, these become conditional transitions
\[
p_\theta\!\big(x^{(n-1)} \mid x^{(n)}, \bm y_{1:t}\big).
\]
Equivalently, one may write the model in terms of a conditional denoiser
$\epsilon_\theta(x^{(n)}, n \mid \bm y_{1:t})$.
Running the reverse chain from the Gaussian prior therefore produces a
sample from the conditional predictive distribution
\[
p_\theta\!\big(\bm y_{t+1:t+H} \mid \bm y_{1:t}\big),
\]
which is the forecasting object introduced in \eqref{eq:diff-pred}.
Other designs may change the noised variables, the reference distribution,
or how conditioning is introduced.

Time series have distinctive structure, including temporal dependence,
seasonality, trend, multiscale patterns, and cross-variable interactions.
Diffusion algorithms are often modified to exploit these features. The models
below illustrate four nonexclusive design choices: generation order, how
conditioning enters, temporal structure within diffusion, and the role of
diffusion in the forecasting system. Table~\ref{tab:diff-models} summarizes
the models discussed below.

\begin{figure}[t]
\centering
\includegraphics[width=0.98\textwidth]{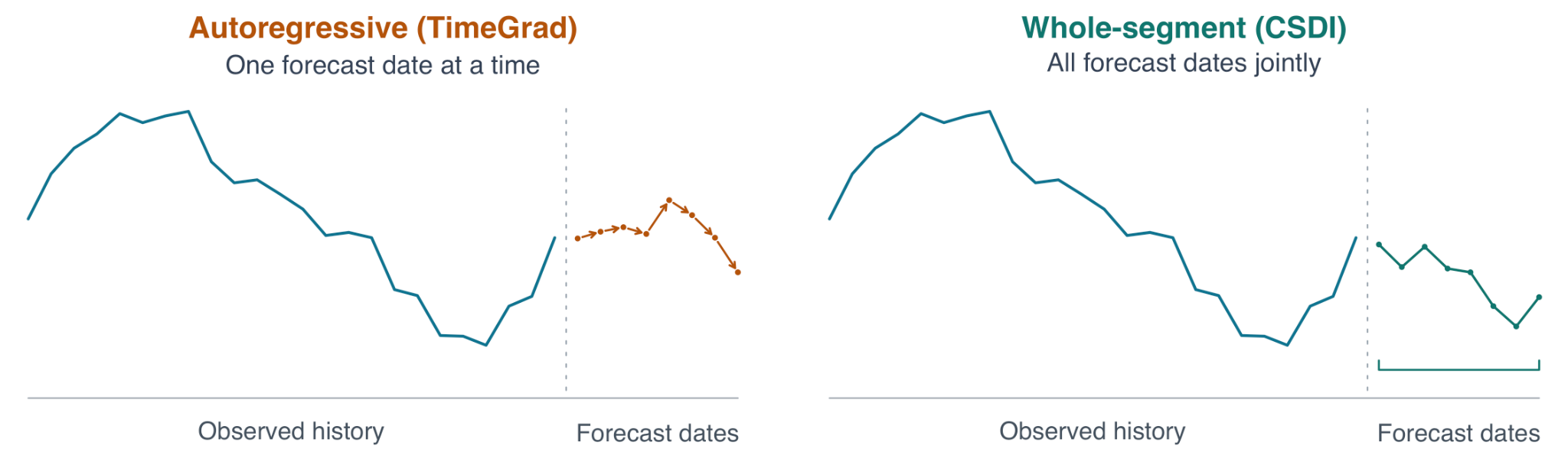}
\caption{Autoregressive versus whole-segment generation.
{In the left panel, TimeGrad samples one future
observation at a time, conditional on the observed history and
any previously sampled values. In the right panel, CSDI
generates the entire future segment by updating all forecast
dates together at each denoising step.} {This schematic illustrates generation order, not relative accuracy
or a guarantee about predictive-band width or forecast drift.}}
\label{fig:ar-vs-joint}
\end{figure}

{The construction above generates the entire future segment
$\bm y_{t+1:t+H}$ in one reverse chain.}
An autoregressive diffusion model instead generates one future observation
vector at a time. At horizon $h$, it conditions on the observed history
and on any future observations already sampled for earlier horizons.
This distinction is related to the choice between iterated and direct
multi-step forecasting \citep{marcellino2006comparison}. Here the question
is how to generate a joint future path.

\par\begingroup\ignorespaces 
For recursive generation, the chain rule gives
\[
p(\bm y_{t+1:t+H}\mid\bm y_{1:t})
=
\prod_{h=1}^{H}
p(\bm y_{t+h}\mid\bm y_{1:t+h-1}).
\]
Sampling successively from the true conditional distributions
therefore gives a draw from the true joint distribution of the
future path. If a fitted conditional distribution is inaccurate,
that error can propagate because later draws depend on earlier
ones. Whole-segment generation avoids this sequential feedback,
although its fitted joint distribution can also be inaccurate.
With either approach, future outcomes can remain uncertain even
under a correctly specified model, and their predictive spread
can vary across forecast horizons.
\par\endgroup 

\citet{rasul2021timegrad} propose TimeGrad, the first diffusion model for
multivariate forecasting. It summarizes the observed history in the hidden
state of a recurrent network. That state is passed to the denoiser as
conditioning information, and the model generates one future date at a time.
{At each date, a full denoising loop produces a multivariate
draw. This draw updates the recurrent state used at the next date
(Figure~\ref{fig:ar-vs-joint}, left).}
\citet{yan2021scoregrad} introduce ScoreGrad, the continuous-time counterpart
of this autoregressive design. A recurrent neural network (RNN) still
summarizes the history, but the conditional distribution at each date is
modelled using a score-based stochastic differential equation (SDE).
A numerical solver for the reverse-time SDE replaces the discrete
denoising loop.
{Errors in the fitted conditional distributions or in the
numerical solution may propagate across forecast dates. These errors
are distinct from genuine predictive variation.}

A different line of work begins with CSDI \citep{tashiro2021csdi}, a
conditional score-based diffusion model originally developed for imputation.
The key observation is that forecasting can be viewed as imputation in which
the missing segment lies at the end of the window. CSDI therefore masks the
future and generates the entire horizon jointly, conditioned on the observed
past.
{Each denoising step updates the masked segment as a whole
(Figure~\ref{fig:ar-vs-joint}, right). The same masking scheme can also be used
to impute missing observations elsewhere in the series.}
\citet{alcaraz2023sssd} propose SSSD, which also treats the future segment
as missing and uses the structured state-space (S4) layers of
\citet{gu2022s4} in its denoiser. These layers capture long-range
dependence at near-linear cost in the sequence length.

A second design choice concerns how conditioning is supplied.
\citet{shen2023timediff} propose TimeDiff, a non-autoregressive forecaster
that strengthens the conditioning signal used by the denoiser. During
training, future mixup blends a representation derived from the observed
history with ground-truth future values. At forecast time, this component
uses only the history-based representation. A second component, called
autoregressive initialization, supplies a linear forecast as additional
conditioning information during both training and forecasting. The reverse
chain itself starts from standard Gaussian noise. The linear forecast
supplies basic temporal patterns, such as short-term trends, while the
non-autoregressive sampler generates the future segment. Both devices improve
performance in the reported experiments.

\citet{kollovieh2023tsdiff} introduce TSDiff, an unconditional diffusion
model for time-series windows. TSDiff is trained without a forecasting
condition. The observed history enters only at sampling time through
self-guidance. The motivation is the decomposition
\begin{equation}
\nabla_x \log p\big(x \mid y_{1:t}\big)
= \nabla_x \log p\big(x\big) + \nabla_x \log p\big(y_{1:t} \mid x\big),
\label{eq:guidance}
\end{equation}
where $x=x^{(n)}$ now denotes the noisy state of a whole window, including
history and future coordinates, at a fixed diffusion step $n>0$. The densities
are understood at that same noise level. The identity shows that an
unconditional score can be converted into a conditional one by adding a
guidance correction. TSDiff approximates this correction at each
denoising step by first using the denoiser to estimate the clean window implied
by the current noisy sample. It then compares the history part of that estimate
with the observed history and nudges the reverse update toward better agreement.
The future part of the generated window gives the forecast.

One trained model can thus be used for forecasting, forecast refinement,
or synthetic generation without retraining. For forecasting, the correction
in \eqref{eq:guidance} is approximated, so the guided sampler is not
guaranteed to draw from the true conditional predictive law. Conditioning
supplied during training targets that law more directly, but ties the model
to the task it was trained for.

A third design choice is how to incorporate temporal scales and
decomposition into the denoising process. Figure~\ref{fig:multiscale}
illustrates two approaches to coarse-to-fine reconstruction. In these
examples, model structure or guidance encourages the recovery of broad
patterns before finer details. Isotropic Gaussian noising alone does
not guarantee this ordering.

\citet{fan2024mgtsd} introduce MG-TSD, which derives coarse-grained versions
of the series by temporal aggregation and uses them as intermediate guidance
targets during diffusion training. These targets retain smoother low-frequency
structure and encourage the reverse process to recover broad temporal patterns
before refining high-frequency detail.
\citet{shen2024mrdiff} propose mr-Diff, which makes the hierarchy explicit
by generating the coarsest trend first and conditioning each finer stage
on the preceding coarser one.
\citet{yuan2024diffts} introduce Diffusion-TS, which places the decomposition
inside the denoiser itself. At each diffusion step, its estimate of the
clean series combines a trend, Fourier-based seasonal components, and a
residual.

\begin{figure}[t]\centering
  \includegraphics[width=0.9\textwidth]{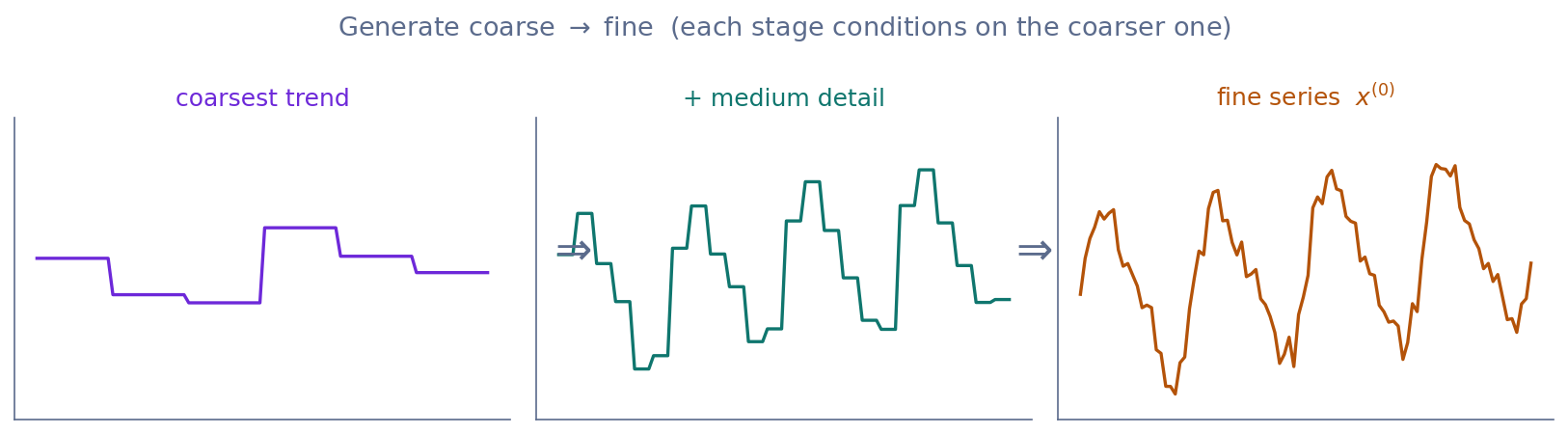}
  \caption{Multiscale structure in diffusion. MG-TSD uses coarse-grained series
as guidance targets at intermediate diffusion levels, whereas mr-Diff explicitly
generates the forecast from coarse trends to finer details.}
  \label{fig:multiscale}
\end{figure}

The preceding models mainly change generation order, conditioning, or temporal
structure within a diffusion sampler. The following two models change the
role of diffusion itself.
\citet{li2022d3vae} propose D$^3$VAE for short and noisy time series, where
whole-segment diffusion models can be difficult to fit with limited data.
In this model, diffusion serves primarily as a denoising and augmentation
device. A coupled forward process perturbs the history and the future
together, creating noisy versions of a small training sample. A
bidirectional variational autoencoder then learns to generate the target
series. The objective includes a denoising term that guides generated
samples toward the observed targets and a penalty that encourages
separation of the latent factors.

\citet{li2024tmdm} propose TMDM, which combines a transformer point forecaster
with a diffusion model for residual uncertainty. The transformer first produces
an estimate of the conditional mean. The forward process diffuses toward a
Gaussian centered at that forecast, $\mathcal{N}(\hat y,I)$, and the same
transformer output also modulates the reverse chain. Diffusion therefore
models plausible deviations around the point forecast. This design connects
the point-forecasting models of Sections~\ref{sec:transformers}--\ref{sec:fm}
with the generative forecasting methods of this section.

\begin{table}[t]\centering
  \caption{Diffusion forecasters, classified by generation scheme, core design,
  and conditioning mechanism. Whole-window generation in TSDiff includes
  history and future coordinates.}
  \label{tab:diff-models}
  \begin{tabular}{llll}
    \toprule
    Model        & Generation      & Core design             & Conditioning       \\
    \midrule
    TimeGrad     & autoregressive  & RNN                     & recurrent state    \\
    ScoreGrad    & autoregressive  & RNN + SDE               & recurrent state    \\
    CSDI         & whole-segment   & attention               & masked observations \\
    SSSD         & whole-segment   & state-space (S4)        & masked observations \\
    TimeDiff     & whole-segment   & convolutional           & future mixup + AR forecast \\
    TSDiff       & whole-window    & unconditional S4        & sampling-time self-guidance \\
    MG-TSD       & autoregressive  & RNN (multi-granularity) & coarse-grained guidance \\
    mr-Diff      & whole-segment   & seasonal--trend         & coarse-to-fine stages \\
    Diffusion-TS & whole-segment   & decomposition attn.     & guidance           \\
    D$^3$VAE     & whole-segment   & diffusion + BVAE        & latent             \\
    TMDM         & whole-segment   & transformer + diff.     & conditional mean   \\
    \bottomrule
  \end{tabular}
\end{table}

\subsection{Evaluation and synthesis}
\label{sec:diff-evaluation}

The models above represent joint uncertainty in different ways. Their
evaluation must assess both marginal forecasts and dependence across
variables and horizons.

\par\begingroup\ignorespaces 
Marginal quantile losses or CRPS assess coordinate-wise predictions, not the
entire dependence structure. For a distribution $P$ on a rescaled future vector
$\bm z$ with finite first moment, the negatively oriented energy score is
\[
\operatorname{ES}(P,\bm z)
=\mathbb E_P\|Z-\bm z\|_2
-\tfrac12\mathbb E_P\|Z-Z'\|_2,
\]
where $Z,Z'$ are independent draws from $P$. For $B\geq2$ independent forecast
paths an unbiased Monte Carlo estimate replaces the first term by the average
distance to the outcome and the second by
$[2B(B-1)]^{-1}\sum_{b\ne b'}\|Z^{(b)}-Z^{(b')}\|_2$.
Scores are minimized, and the scales used to compare variables and horizons
must be fixed from training data. Proper joint scores should be complemented
by coverage and economically relevant joint-event diagnostics
\citep{gneiting2007strictly}. A cloud of simulated paths demonstrates the
fitted model's uncertainty representation, not calibration, parameter-uncertainty
coverage, or structural validity. The empirical illustration below evaluates
point forecasts only and makes no joint-probabilistic performance claim.
\par\endgroup 

\par\begingroup\ignorespaces 

Diffusion's principal contribution here is flexible joint
probabilistic forecasting
\citep{rasul2021timegrad,tashiro2021csdi}.
In a whole-path specification, the model represents a conditional
distribution for the $pH$-dimensional future vector, allowing
cross-variable and cross-horizon dependence to be represented
jointly. This ability to generate a high-dimensional predictive
distribution should be distinguished from reducing the statistical
complexity of estimating it. The diffusion mechanism does not
itself provide dimension reduction or guarantee reliable
high-dimensional estimation.
Shared denoising parameters and structured representations can
exploit common temporal and cross-variable patterns. Latent
representations and regularization, where used, provide additional
ways to control effective model complexity. Moreover, the absence
of an explicit dimension-reduction step does not preclude
adaptation to underlying low-dimensional structure.
\citet{chen2023score} establish score-estimation and
distribution-recovery guarantees for data supported on an unknown
low-dimensional linear subspace, using an appropriately structured
score network. \citet{tang2024adaptivity} establish adaptation to
smooth low-dimensional manifold structures. Under their respective
structural and regularity assumptions and suitable network
specifications and tuning, the principal convergence-rate exponents
are governed by intrinsic rather than ambient dimension.

These results establish the possibility of exploiting
low-dimensional structure; they do not provide a general guarantee
for conditional forecasting with dependent time-series observations.
We therefore distinguish the dimension of the forecast target,
the complexity of the fitted model, and its possible adaptation
to intrinsic structure. Generating joint future paths alone does
not establish accurate estimation from a short sample.
\par\endgroup


For nonstationarity, the picture is less settled. Recent work has begun to
incorporate time-series characteristics into diffusion design. For example, \citet{lee2024ant} select a dataset-specific noise schedule
using autocorrelation statistics that summarize temporal dependence. This remains an adaptive scheduling device rather than a
formal treatment of unit roots or cointegration. The diffusion designs reviewed here, like the foundation models of
Section~\ref{sec:fm}, do not enforce cointegration or long-run equilibrium
by construction. Where such structure must hold, the vector error-correction
model in \eqref{eq:vecm} remains the appropriate benchmark.
{Diffusion provides a flexible implicit family that can approximate multimodal, asymmetric, or heavy-tailed predictive shapes when its architecture, objective, training data, and sampler support them. This is greater flexibility than a specified Gaussian law, not freedom from distributional restrictions or a general tail-calibration guarantee.}The broader
trajectory is to place such generative heads on top of pretrained backbones,
combining transformer representations, pretraining-based generalization, and
diffusion-based scenario uncertainty.

\newpage
\section{Synthesis and outlook}
\label{sec:synthesis}

\par\begingroup\ignorespaces 
Forecasting methods should be compared as complete specifications.
For example, a transformer can be fitted with or without a LASSO-type
penalty on the weights of its forecasting layer. Architecture determines how the forecasting
function is constructed, and regularization controls its estimation.
Pretraining determines the additional information used to learn
representations, and the forecasting layer or generative specification
determines the predictive output. These choices can be combined.

Transformers address nonlinearity through input-dependent attention
weights and can represent the observed history using different types
of tokens. They use shared parameters to compute these weights and may
also employ sparsity-inducing penalties or other forms of shrinkage
to control model complexity. Regime models allow the dynamics to vary
across states. Estimating those dynamics in a high-dimensional system
may still require regularization, and a process that switches between
regimes can be stationary.

{Likewise, representing a joint future-path law with
diffusion does not itself provide dimension reduction.
Diffusion models may nevertheless adapt to intrinsic
low-dimensional structure under suitable assumptions and
model choices, as discussed in
Section~\ref{sec:diff-evaluation}. This possibility should be
distinguished from a general guarantee of reliable
high-dimensional estimation.}

Table~\ref{tab:synth} separates these contributions. Comparisons
should hold the forecasting target and admissible information set
fixed where possible, disclose external training data, and account
for tuning, training, and forecast-time costs. The purpose is to
identify where attention, pretraining, or diffusion adds value
relative to strong alternatives, not to establish an unconditional
ranking of model families.
\par\endgroup

These advances make the modern architectures flexible representations of the conditional law of the
future given the past or selected functionals of that law. Flexibility is not by itself an econometric standard. For a method to inform
policy, risk management, or scientific interpretation, its use must also be governed by credible
evaluation, calibrated uncertainty where required, and a clear account of the
nonstationarity it is meant to handle. Structural causal claims additionally
require explicit identification.

\par\begingroup\ignorespaces 

Diffusion
samples describe variation in future paths conditional on the fitted model.
Accounting for uncertainty in the fitted parameters requires an additional
inferential step.
\par\endgroup 

\par\begingroup\ignorespaces 
Increasing the number of simulated paths reduces Monte Carlo error for the
fitted law, not uncertainty about that law or the need to assess calibration.
Calibration should be assessed on held-out observations using interval coverage
and distributional diagnostics at the relevant horizons. Conformal guarantees
must specify their coverage target and assumptions. Exchangeability-based
finite-sample coverage does not extend automatically to an arbitrary dependent
or changing time series.
\par\endgroup

Evaluation should mimic the information set available at the forecast origin. {The chronological designs in Appendix~\ref{sec:practice}
implement this principle for estimation and tuning. Alternative validation
schemes require their own justification under the dependence and prediction target.} For
pretrained models a further concern is contamination. {A nominally zero-shot score cannot establish transfer to unseen outcomes when those outcomes overlap the pretraining data. Verified post-cutoff targets or documented corpus separation strengthen the interpretation; private ownership alone does not certify separation.} Modern models should be compared against strong classical baselines, including
seasonal naive forecasts, ARIMA, VAR or Bayesian VAR, and a VECM where cointegration matters.

\par\begingroup\ignorespaces 
In the exploratory comparison in Table~\ref{tab:macro-patch-rmse}, the
selected patched specification lowers inflation and production-growth RMSEs
relative to the unpatched transformer, but raises interest-rate RMSEs and
the aggregate relative RMSE. These descriptive differences, obtained on a
reused evaluation period, do not isolate the effect of patching or establish
an overall ranking of forecasting methods.
\par\endgroup 

\par\begingroup\ignorespaces 
Structural use requires the identifying assumptions discussed in
Section~\ref{sec:classical}. In the macroeconomic example, a simulated
interest-rate increase is not automatically a monetary-policy intervention,
and predictive relevance of past rates does not identify the effect of
changing policy. The comparison in Appendix~\ref{sec:application} evaluates
point-forecast errors. It does not identify policy effects, test Granger
noncausality, or assess predictive-interval coverage.
\par\endgroup 

Nonstationarity is not a single problem. Series may carry stochastic or deterministic trends,
structural breaks, evolving volatility, or gradual parameter drift, and classical econometrics
distinguishes these cases because each calls for a different response. Modern systems tend to
address them together, through scaling, normalization, long context windows, or pretraining across
regimes. These devices can improve robustness but do not replace an explicit account of the
instability. When series are cointegrated, differencing without retaining
error-correction structure can discard long-run equilibrium information.
Normalization has this risk when relevant level and scale information is
discarded, rather than retained or restored.

Taken together, the move from vector autoregressions to transformers, foundation models, and
diffusion is best understood as an expansion of the forecasting class, not a replacement of
econometric structure. {Self-attention lets the observed window determine how
much weight each transformed token receives.} Pretraining across many series enables
frozen zero-shot forecasting and can also complement per-data-set estimation.
Diffusion provides flexible mechanisms for sampling joint future paths from a
fitted predictive law. These developments expand flexibility, scale, and the
available predictive outputs. Structural interpretation and causal policy analysis
still require the restrictions, identification, and inferential justification
appropriate to the question. In particular, the burden of dimensionality is partly shifted from estimating many
parameters on one data set to learning shared representations from large and heterogeneous corpora.
{Shared representations impose architectural restrictions and can be combined
with explicit regularization. Pooling and pretraining additionally change the
information available to learn these representations; they do not remove the
need to control effective complexity.}

The task ahead is to combine the predictive strength of modern architectures with the structural
and inferential tools of econometrics. Developing reliable inference and interpretability for these
models is a promising direction, because it would clarify when their predictive distributions are
calibrated, when learned restrictions can be interpreted econometrically, and how far their
black-box representations can support policy or scientific claims rather than forecasts alone.
{Determining which combinations of architecture, regularization, and training
information yield reliable forecasting improvements under dependence,
heterogeneity, and instability remains, in our view, a central open problem.}

\par\begingroup\ignorespaces 
\appendix

\par\begingroup\ignorespaces 
\section{A practical forecasting workflow}

\label{sec:workflow}
\label{sec:practice}

The preceding comparisons suggest a common workflow for
implementing and evaluating forecasting models. The starting
point is the forecasting task and the available information.
Model complexity, pretraining, and generative sampling are
useful only if they improve the required forecasts at an
acceptable computational cost.

\begin{enumerate}[leftmargin=*]

\item \textbf{Define the forecasting task.}
Specify the variables, sampling frequency, forecast horizons,
and information available at each forecast date. Decide whether
the required output is a point forecast, marginal quantiles or
distributions, or joint future paths. Record publication delays,
data revisions, and missing observations. Future covariates
may be included only if they are known when the forecast is made.

\item \textbf{Choose relevant comparisons.}
Table~\ref{tab:forecast-workflow} links common application
features to candidate methods and practical checks. Include
simple baselines, such as naive or seasonal-naive forecasts,
univariate autoregressions, and a VAR with suitable
regularization. Add nonlinear or pretrained models according
to the task. Use comparable information sets where possible
and report differences when a model cannot accept the same
inputs. The table identifies starting points for comparison;
the final choice depends on validation performance.

\item \textbf{Select settings using earlier data.}
Divide the observations chronologically into training,
validation, and test periods. Use rolling forecasts within
the validation period to select lag or context lengths,
regularization, model versions, and other settings.
Choose an expanding or rolling estimation window, including its length
when applicable.
Fit preprocessing within each training window.
Choose any calibration procedure, parameter-update schedule,
and computing budget before evaluating the test period.

For neural models, record the training objective, optimization
settings, and stopping rule. For pretrained models, record the
model version, documented training period, and any known overlap
with the evaluation data. If overlap cannot be checked, report
the contamination status as unverified. For diffusion models,
also specify the sampling method, number of denoising steps,
and number of generated paths.

\item \textbf{Produce successive forecasts.}
At each test forecast date, use only the information available
at that date. Training examples must have all their target
observations already observed. Fit or refit application-specific
models before the first test forecast, then update them according
to the chosen schedule. Fit or update any data-dependent external
preprocessing using permitted past data under its selected rule,
including for zero-shot models. Retain fitted preprocessing between
prescribed updates; any normalization calculated separately for each
input window follows the model's specified rule.

For zero-shot forecasting, keep the pretrained parameters fixed
and supply the new observed window. Evaluate fine-tuning as a
separate procedure. If calibration is used, apply the selected
rule using only past forecast errors whose outcomes are already
observed. Save each forecast before its outcomes become available.

\item \textbf{Evaluate the required output.}
Match the forecast summary to the loss: squared error for mean
forecasts and absolute error for median forecasts. Use quantile
loss for quantiles or the continuous ranked probability score
(CRPS) for marginal predictive distributions. Report interval
coverage together with width. For joint future paths, add a
joint score, such as the energy score, and checks of relevant
path features. Marginal coverage alone does not assess the
dependence across variables and horizons.

Compare models on the same forecast dates and horizons.
Report results separately by variable and horizon, and state
how losses are aggregated. Any scaling used to combine errors
across variables should be determined from training data.

\item \textbf{Report resources and preserve the experiment.}
Report fitting time, forecasting time, hardware, and memory
requirements. For sampling-based forecasts, include the number
of generated paths. Record failed or missing forecasts rather
than silently excluding them. Save data versions, transformations,
forecast dates, model settings, software versions, random seeds,
and forecasts so that the reported comparisons can be reproduced.

\end{enumerate}

\begin{table}[!htbp]
\centering

\small
\renewcommand{\arraystretch}{1.12}
\caption{A decision guide for an initial forecasting comparison.
Candidate methods should be assessed using the same forecasting
task and an appropriate validation design.}
\label{tab:forecast-workflow}
\label{tab:decision}
\begin{tabular}{@{}>{\raggedright\arraybackslash}p{0.22\textwidth}
                  >{\raggedright\arraybackslash}p{0.35\textwidth}
                  >{\raggedright\arraybackslash}p{0.35\textwidth}@{}}
\toprule
Application feature & Candidate methods & Main practical check \\
\midrule
Short history and a modest number of variables &
Univariate AR, Bayesian VAR, and a frozen pretrained model &
Sensitivity to shrinkage and whether pretraining transfers
to the application \\
\addlinespace
Many related predictors &
Sparse or factor VAR, regularized regression, pooled trees,
and neural models &
Whether additional predictors improve forecasts after
accounting for model complexity \\
\addlinespace
Nonlinear or regime-dependent dynamics &
Threshold or switching models, trees, and attention models &
Sufficient observations across relevant states and performance
in different periods \\
\addlinespace
Integrated variables with an equilibrium relation &
VECM or a flexible model incorporating the relation &
Integration order, cointegrating rank, and long-horizon behaviour \\
\addlinespace
Marginal predictive uncertainty &
Models with quantile outputs or predictive distributions &
Quantile loss or CRPS, together with interval coverage and width \\
\addlinespace
Joint future paths &
VAR path simulation, distributional neural models, and diffusion &
Dependence across variables and horizons, joint scores,
and sampling cost \\
\addlinespace
Limited forecasting time &
Parsimonious statistical models and compact pretrained models &
Total forecasting time, including preprocessing and sampling \\
\bottomrule
\end{tabular}
\end{table}

Table~\ref{tab:forecast-protocol} summarizes the procedure.
The same forecasting sequence is used during validation and
testing, but model settings are selected before the test period.

\begin{table}[!htbp]
\centering
\small
\renewcommand{\arraystretch}{1.12}
\setlength{\tabcolsep}{6pt}
\caption{A chronological forecasting and evaluation protocol.}
\label{tab:forecast-protocol}

\begin{tabularx}{\linewidth}{
  @{}>{\raggedright\arraybackslash}p{0.20\linewidth}
     >{\raggedright\arraybackslash}X@{}
}
\toprule
\textbf{Step} & \textbf{Procedure} \\
\midrule

\multicolumn{2}{@{}l}{\textbf{1. Set up the experiment}} \\
\addlinespace[3pt]

Define the task &
Specify targets, forecast horizons, available inputs, and
candidate models. Choose chronological training, validation,
and test periods. \\

Select settings &
Use rolling validation to select model settings, preprocessing
rules, calibration procedures if used, expanding or rolling
estimation windows, update schedules, and computing budgets. \\

Document pretraining &
Load and document the selected pretrained model versions. \\

\midrule
\multicolumn{2}{@{}l}{
  \textbf{2. At each test forecast date \(t\)}
} \\
\addlinespace[3pt]

Establish information &
Construct the information set available before forecasting.
Carry out the following steps for each candidate model. \\

Update preprocessing &
At the first test date or when a preprocessing update is due,
fit the required preprocessing using permitted past data.
Otherwise, retain the fitted preprocessing. \\

Update the model &
For zero-shot models, keep pretrained parameters fixed.
For other models, fit at the first test date and at scheduled
updates, using training examples whose required outcomes are
already available. Otherwise, retain the fitted model. \\

Generate forecasts &
Prepare inputs using the specified preprocessing rule.
Produce forecasts for \(y_t,\ldots,y_{t+H-1}\) on the target
scale. \\

Calibrate and save &
If calibration is used, apply the selected procedure using
past forecast errors whose outcomes are already observed.
Save forecasts, relevant predictive samples, and runtime. \\

\midrule
\multicolumn{2}{@{}l}{
  \textbf{3. Evaluate as outcomes become available}
} \\
\addlinespace[3pt]

Assess performance &
Score models on matched forecast dates and horizons.
Assess uncertainty and joint-path quality where applicable. \\

Report and archive &
Report forecast performance and computing costs.
Archive the data, settings, forecasts, and evaluation code. \\

\bottomrule
\end{tabularx}
\end{table}

Appendix~\ref{sec:application} illustrates the workflow with a revised-data
pseudo-out-of-sample comparison of point forecasts.

\par
\par\endgroup 
\par\begingroup\ignorespaces 
\section{A reproducible macroeconomic forecasting illustration}

\label{sec:application}

We implement the running example using monthly US consumer prices
(\texttt{CPIAUCSL}), industrial production (\texttt{INDPRO}), and the
federal funds rate (\texttt{FEDFUNDS}) from FRED.\footnote{ Source documentation:
\url{https://fred.stlouisfed.org/series/CPIAUCSL},
\url{https://fred.stlouisfed.org/series/INDPRO}, and
\url{https://fred.stlouisfed.org/series/FEDFUNDS}.
The replication archive contains the retrieved source values, transformations,
checksums, and source metadata; it does not rely on a changing live download.}
The price and production indices are seasonally adjusted. For January 1960
through December 2024, define
\[
\bm y_t=(\pi_t,g_t,r_t)^\top,\qquad
\pi_t=1200\Delta\log P_t,\qquad g_t=1200\Delta\log I_t,
\]
with $r_t$ in percent. The first two variables are annualized monthly log
changes, not year-on-year rates; industrial production is an output proxy,
not GDP. One additional raw observation, December 1959, initializes the
transformations. No missing values are imputed and no pandemic observations
are removed. The frozen revised-data tables were retrieved on 6 November 2026;
the displayed source updates were 14 July, 17 July, and 3 August 2026,
respectively. They are not asserted to be the latest live release or a
certified common-date vintage. This is a revised-data pseudo-out-of-sample
exercise, not a reconstruction of historical data vintages or release-time
information sets.

We forecast the individual monthly vector $\bm y_{t+h}$, not cumulative
$h$-month growth, at $h\in\{1,3,6,12\}$. Candidate models are initially
estimated through December 2014 and validated at 49 common origins from
December 2014 through December 2018, with all validation targets observed
by December 2019. Selection minimizes squared errors standardized by each
variable's initial-training standard deviation, averaged across origins,
variables, and horizons. Hyperparameters are then fixed. Test models are
refitted on expanding samples each December from 2019 to 2023 and held fixed
for the following 12 origins, while the observed history updates monthly.
Scaling uses the fit sample only. Direct models use only complete target
blocks observed by the fit cutoff. Targets end in December 2024, giving
60, 58, 55, and 49 matched origins at the four horizons. Models share origins
within each horizon, but the origin counts differ across horizons.

Direct LASSO and the unpatched transformer forecast changes from the
latest observed value and add that value back. Persistence, which
forecasts the latest observation at every horizon, provides the reference
for the relative-error summaries. Validation selects a 24-month input
and penalty of 0.1 for direct LASSO from lag lengths $\{6,12,24\}$ and
penalties $\{0.001,0.01,0.1\}$. The complete benchmark settings and
results are retained in the replication archive.

The unpatched transformer uses 24 multivariate time-point tokens,
width 16, two heads, one attention block, a feed-forward width of 32,
learned positions, dropout 0.1, and a linear forecasting layer on the
final token. Each network has 2,908 parameters. Training uses Adam at
learning rate 0.001, full-batch standardized squared loss, and gradient
clipping. Validation selects 80 epochs from $\{40,80,120\}$. Forecasts
average predictions from three fixed seeds, whose individual results
are also archived. No external pretraining is used. This architecture
is fixed; its validation search varies training duration.

To examine sensitivity to the unpatched architecture, we conduct
an exploratory extension on the same data and forecast origins. The
structured search contains 48 specifications and four training durations
(40, 80, 120, and 200 epochs), giving 192 configurations evaluated using
three-seed forecast averages. It varies patch lengths of 1, 3, 6, and
12 months, contexts of 24 or 48 months, one, two, or four heads, one or
two blocks, and widths of 16 or 32, alongside output aggregation and
normalization choices. The grid includes joint-variable patches,
channel-independent patches, variable tokens, unpatched controls, and
a feed-forward control; it is not a full factorial design. The patch and
variable-token models are compact implementations inspired by
\citet{nie2023patchtst} and \citet{liu2024itransformer}, not reproductions
of their complete benchmark configurations. Selection uses the same
pre-2020 standardized mean-squared-error criterion. The original
2020--2024 results had already been inspected before this extension;
although new test scores do not select architectures, epochs, or seeds,
the reused period is not a fresh confirmatory holdout.

Validation selects a channel-independent transformer with a 24-month
context split into two nonoverlapping 12-month patches per variable,
embedding width 16, two attention heads, one block, and feed-forward
width 32. The embedding, encoder, and forecasting layer are shared across
variables. Both patch representations enter the forecasting layer,
which predicts standardized levels directly at the four horizons,
without automatically adding the latest observation. The model has
2,628 parameters, uses 200 epochs, and averages the same three seeds
as the unpatched model. There is no external pretraining or $\ell_1$
penalty. Its validation loss is 0.255, compared with 0.303 for the
unpatched transformer. Validation minimizes standardized mean squared error,
whereas the reported aggregate averages RMSE ratios to persistence; the two
criteria need not rank models identically. The full grid and all twelve evaluated finalists
and controls are documented in the supplementary replication report.
\par\endgroup 

\begin{table}[t]
\centering
\par\begingroup\ignorespaces 
\small
\caption{Out-of-sample RMSE for direct LASSO and two transformer
specifications. Individual-horizon results use targets in 2020--2024;
aggregate scores are also reported for 2021--2024. The patched
transformer is selected using pre-2020 validation. Bold entries identify
the lowest unrounded value among the three displayed specifications
in each row; comparisons are descriptive.}
\label{tab:macro-patch-rmse}
\setlength{\tabcolsep}{8pt}
\begin{tabular}{@{}lrrr@{}}
\toprule
Horizon & Direct LASSO & Unpatched transformer & Patched transformer \\
\midrule
\multicolumn{4}{l}{\textit{Inflation}} \\
1 & 3.367 & 3.675 & \textbf{3.333} \\
3 & 4.002 & 4.666 & \textbf{3.693} \\
6 & 3.630 & 4.519 & \textbf{3.540} \\
12 & \textbf{4.153} & 5.223 & 4.196 \\
\addlinespace
\multicolumn{4}{l}{\textit{Industrial-production growth}} \\
1 & 27.988 & 33.537 & \textbf{27.083} \\
3 & 31.575 & 41.450 & \textbf{27.827} \\
6 & 16.990 & 29.652 & \textbf{15.967} \\
12 & 11.170 & 29.283 & \textbf{11.060} \\
\addlinespace
\multicolumn{4}{l}{\textit{Interest rate}} \\
1 & \textbf{0.248} & 0.249 & 0.627 \\
3 & 0.654 & \textbf{0.604} & 1.042 \\
6 & 1.157 & \textbf{0.972} & 1.444 \\
12 & 1.907 & \textbf{1.685} & 2.226 \\
\addlinespace
\midrule
\multicolumn{4}{l}{\textit{Average relative RMSE across all twelve cells}} \\
2020--2024 & \textbf{0.795} & 0.926 & 0.987 \\
2021--2024 & \textbf{0.821} & 0.896 & 1.052 \\
\bottomrule
\end{tabular}
\par\smallskip
\begin{minipage}{0.97\textwidth}\footnotesize
Inflation and production growth are annualized monthly log changes;
interest rates are in percent. Individual-horizon RMSEs are measured
in percentage points of these variables. For 2020--2024, the numbers
of matched origins are 60, 58, 55, and 49 at horizons 1, 3, 6, and 12,
respectively. The final two rows are equally weighted averages of the
twelve variable--horizon RMSE ratios to persistence and are
dimensionless. The 2021--2024 row rescores the same forecasts without
retuning or removing 2020 observations from subsequent training
samples. Neural forecasts average three fixed seeds.
\end{minipage}
\par\endgroup 
\end{table}

\par\begingroup\ignorespaces 
Table~\ref{tab:macro-patch-rmse} shows that the selected patched
transformer achieves lower RMSE than direct LASSO for inflation at
horizons 1, 3, and 6 and for production growth at all four horizons:
seven of these eight comparisons. It also improves on the unpatched
transformer in all eight inflation and production-growth comparisons.
These results illustrate the gains that a different transformer
configuration can deliver for particular forecasting targets.

Performance nevertheless varies across variables. The patched model
has higher interest-rate RMSE than both comparators at all four horizons.
Its average relative RMSE across all twelve variable--horizon
combinations is therefore 0.987, compared with 0.926 for the unpatched
transformer and 0.795 for direct LASSO. On 2021--2024 targets the
corresponding scores are 1.052, 0.896, and 0.821. Direct LASSO thus
retains the lowest aggregate error among the three displayed models.
Since patching, channel processing, the forecasting layer, and training
duration change together, the gains cannot be attributed to patch size
alone.

The original archive supplies frozen source values, transformations,
settings, software versions, fitted models, and all 7,992 forecast
records, together with tables, figures, and replay checks. These
original results are unchanged. The separate patch/structure archive
adds the locked grid and selections, all validation scores, forecasts
for every evaluated finalist, and fitted networks. Its audit records
replay 31,968 forecast records from 180 saved networks and independently
rescore all 192 validation configurations. The original and extended
experiments are kept separate, including their selection records and
limitations. These checks establish internal reproducibility, not
historical release-time availability, statistical significance, or
independent verification against a different data vintage.
\par\endgroup 

\section{Reproducibility of the GIFT-Eval comparison}
\label{app:gifteval}
\label{app:gift-audit}

The original access date for Table~\ref{tab:gifteval} is
14 July 2026. For reproducibility, we checked its displayed
values against the following version of the
SalesforceAIResearch/gift-eval repository:
\begin{quote}
\small
\url{https://github.com/SalesforceAIResearch/gift-eval/tree/6fdb10df9c17411f0aef5ff862afbec23627c12f}
\end{quote}
This commit is dated 6 July 2026 and reproduces every
reported value to three decimal places. The original download log is not
available; this identifies the reconstructed snapshot. The archived source
files are hashed in \texttt{benchmark\_audit/manifest.json}, and the
accompanying script regenerates the aggregate scores without fitting the models.

Each directory below contains the result file
\path{all_results.csv} and its accompanying
\path{config.json}. Paths are relative to the pinned
repository version.

\begin{center}
\small
\begin{tabular}{@{}ll@{}}
\toprule
Table entry & Result directory \\
\midrule
Seasonal naive & \path{results/seasonal_naive/} \\
Auto-ARIMA     & \path{results/auto_arima/} \\
DeepAR         & \path{results/deepar/} \\
DLinear        & \path{results/DLinear/} \\
N-BEATS        & \path{results/N-BEATS/} \\
Moirai-2.0     & \path{results/Moirai2/} \\
TimesFM-2.5    & \path{results/TimesFM-2.5/} \\
Chronos-2      & \path{results/chronos-2/} \\
\bottomrule
\end{tabular}
\end{center}

The calculation uses the column
\path{eval_metrics/mean_weighted_sum_quantile_loss}.
Let $\ell_{m,d}$ denote this loss for model $m$ on task $d$,
and let $\ell_{0,d}$ denote the seasonal-naive loss.
The reported aggregate is
\[
R_m
=
\exp\left\{
\frac{1}{97}\sum_{d\in\mathcal D}
\log\left(\frac{\ell_{m,d}}{\ell_{0,d}}\right)
\right\},
\qquad |\mathcal D|=97.
\]
Tasks are matched by their dataset--frequency--horizon
identifiers. All eight files contain the same 97 tasks
and positive, finite losses. No task scores were excluded
or imputed in this aggregation.

Both blocks report $R_m$. A percentage skill defined as $100(1-R_m)$
would be a transformation of this aggregate, not a universal conversion
from another source's skill convention. These marginal quantile scores
do not assess the complete joint future-path law.

This does not rule out substitutions made when the benchmark
results were generated. The statistical-model protocol in
\citet[Appendix A]{aksu2024gifteval} replaces results from runs
exceeding its time limit with seasonal-naive results.
The archived task-level files do not identify which results
used this fallback.

The archived configurations identify the pretrained models as
\begin{itemize}
\item Moirai-2.0:
\path{Salesforce/moirai-2.0-R-small};
\item TimesFM-2.5:
\path{google/timesfm-2.5-200m-pytorch};
\item Chronos-2:
\path{amazon/chronos-2}.
\end{itemize}
The associated evaluation implementations are
\path{notebooks/moirai2.ipynb},
\path{notebooks/timesfm2p5.ipynb}, and
\path{notebooks/chronos-2.ipynb} at the same repository
version. They use model-produced quantiles at levels
$0.1,0.2,\ldots,0.9$ for the quantile-loss calculation.
The configurations and notebooks do not pin immutable
model-weight revisions.

For the task-specific neural baselines,
\citet[Appendix A]{aksu2024gifteval} describe using
probabilistic output heads where feasible and otherwise
representing a point forecast by a single predictive sample.
The archived configurations do not identify every model's
output construction or fitted distribution.
Consequently, the table should be interpreted as a comparison
of the published benchmark implementations, rather than
assuming that every baseline supplies a nondegenerate
predictive distribution.

The three pretrained-model configurations report no test-data
leakage. The model documentation provides more specific
information: Moirai-2.0 includes historical observations
from GIFT-Eval training splits; Chronos-2 excludes the test
portions of overlapping source datasets; and TimesFM-2.5
lists GIFT-Eval Pretrain together with other sources and
source-specific cutoffs.\footnote{Checkpoint documentation:
\url{https://huggingface.co/Salesforce/moirai-2.0-R-small};
\url{https://huggingface.co/amazon/chronos-2};
\url{https://huggingface.co/google/timesfm-2.5-200m-pytorch}.}

These statements document the reported separation of training
and test observations. They do not establish that every
evaluation series was entirely absent from pretraining,
or constitute an independent audit of all training data.
The archived scores permit reproduction of the table's
aggregates; exact reproduction of the underlying model runs
also requires the weight revisions and implementation details
not recorded in the archive.

The example notebooks do not establish the exact historical run configuration:
for example, the TimesFM notebook defaults to a demonstration subset even
though its submitted CSV contains all 97 tasks. Our separate public data
experiment in Appendix~\ref{sec:application} supplies fitted models, forecast
records, settings, and software details, and uses no external pretraining.
\par\endgroup 

\section{Summary tables}
\begin{table}[p]
\centering
\par\begingroup\ignorespaces 
\small
\caption{Forecasting approaches and their assumptions, inputs, outputs, and
computation. The categories follow Section~\ref{sec:intermediate}.
Transformers are a neural-network architecture within general nonlinear
machine-learning models. Pretraining is a training choice rather than a
separate model class. Diffusion specifies a generative predictive model.
Resource requirements depend on dimensions, tuning, and implementation.}
\label{tab:bridge}
\setlength{\tabcolsep}{4pt}
\renewcommand{\arraystretch}{1.12}
\begin{tabularx}{\textwidth}{@{}>{\raggedright\arraybackslash}p{3.0cm}
*{3}{>{\raggedright\arraybackslash}X}@{}}
\toprule
Approach & Main restrictions and data requirements & Covariates and outputs
& Main computational burden \\
\midrule
LASSO-type methods
& Exact or approximate sparsity; dependence and regressor conditions;
  penalty and lag tuning
& Available covariates; point forecasts; VAR paths require an innovation law
& Penalized estimation and tuning \\
\addlinespace
Bayesian VARs
& Prior centre and tightness; shrinkage across variables and lags
& VARX inputs; posterior predictive distributions
& Posterior estimation and predictive simulation \\
\addlinespace
Factor models
& Low-rank common component; strong factors; restricted idiosyncratic dependence
& Model-specific covariates; factor-based forecasts
& Factor estimation and forecasting \\
\addlinespace
Structural breaks and parameter drift
& Break magnitudes and within-segment information, or a regularized
  parameter-evolution law
& Model-specific inputs; forecasts with changing parameters
& Break search, filtering, or posterior simulation \\
\addlinespace
Threshold and switching models
& Threshold or regime assumptions; sufficient data within regimes;
  regularization in large systems
& Model-specific covariates; state-dependent forecasts
& Threshold search, filtering, or posterior simulation \\
\addlinespace
General nonlinear machine-learning models
& Feature construction, model complexity, regularization, and tuning;
  includes trees, kernels, and neural networks
& Available covariates; output depends on the model and loss
& Feature construction, fitting, and tuning \\
\addlinespace
\hspace*{0.4em}Transformers
& Tokenization and shared attention projections; optional penalties;
  sufficient training information
& Architecture-dependent covariates; point, quantile, or distributional output
& Network training; dense attention quadratic in token count \\
Diffusion forecasters
& Conditional generative structure; adequate training information
& Supported covariates; joint samples when dependence is modelled
& Denoiser training and repeated denoising per sample \\
\bottomrule
\end{tabularx}
\par\endgroup 
\end{table}

\begin{table}[p]
\centering
\par\begingroup\ignorespaces 
\small
\caption{Contributions to the three forecasting challenges.
The modelling choices in different rows can be combined;
no entry is an automatic statistical or forecasting guarantee.}
\label{tab:synth}
\setlength{\tabcolsep}{3pt}
\renewcommand{\arraystretch}{1.12}
\begin{tabularx}{\textwidth}{@{}>{\raggedright\arraybackslash}p{3.0cm}
*{3}{>{\raggedright\arraybackslash}X}@{}}
\toprule
Specification & High dimensionality & Nonstationarity
& Nonlinearity / predictive law \\
\midrule
VAR and VECM
& Sparsity, shrinkage, or factor structure where imposed
& Integration and cointegration where specified
& Linear mean; innovations need not be Gaussian \\
\addlinespace
Threshold / switching
& Additional restrictions required in large regime-specific systems
& Recurrent switching can be stationary
& State-dependent dynamics and predictive laws \\
\addlinespace
Structural breaks and parameter drift
& Complexity depends on break or evolution restrictions and regularization
& Abrupt breaks or parameter drift
& Changing dynamics; linear specifications possible \\
\addlinespace
General nonlinear machine-learning models
& Feature selection and model-specific complexity controls
& Updating or window selection; no automatic equilibrium restriction
& Nonlinear prediction; outputs depend on the model and loss \\
\addlinespace
\hspace*{0.4em}Transformers
& Parameter sharing plus optional regularization, including LASSO-type penalties
& {Context adaptation; no automatic treatment of breaks or cointegration}
& Nonlinear attention and feed-forward maps; output depends on layer and loss \\
\addlinespace
Diffusion forecasters
& {No automatic dimension
reduction; possible adaptation to intrinsic low-dimensional
structure under suitable assumptions. Complexity depends
on the denoising architecture and optional regularization}
& Relevant instability needs appropriate conditioning or constraints
& Flexible implicit predictive laws; dependence and calibration require evaluation \\
\bottomrule
\end{tabularx}
\par\endgroup 
\end{table}

\par\begingroup\ignorespaces 

Table~\ref{tab:synth} compares modelling contributions.
{Table~\ref{tab:inference} distinguishes inference about
estimated parameters from uncertainty about future observations, and
separately considers predictive relevance and structural identification.
In all model families, interpreting a response as the effect of an economic
shock requires identifying assumptions.}
\par\endgroup 

\begin{table}[t]
\centering
\par\begingroup\ignorespaces 
\small
\caption{{\textbf{Uncertainty and inference}}. Both econometric and flexible predictive
models require assumptions and procedures appropriate to the object of interest.}
\label{tab:inference}
\setlength{\tabcolsep}{4pt}
\renewcommand{\arraystretch}{1.16}
\begin{tabularx}{\textwidth}{@{}>{\raggedright\arraybackslash}p{2.65cm}
*{2}{>{\raggedright\arraybackslash}X}@{}}
\toprule
Object & Econometric benchmark & Flexible predictive model \\
\midrule
Parameter inference
& {Parameter inference under standard assumptions}, for identified parameters or functionals;
  regularization or drift can change the theory.
& Specify an identifiable parameter or functional and justify the inference
  procedure for the learner and dependence structure; fitting alone is insufficient. \\
\addlinespace
Predictive uncertainty
& An innovation law generates forecast variation; parameter uncertainty is
  included only when integrated or otherwise accounted for.
& {With quantile outputs: Forecast intervals from fitted quantiles.
Foundation models: Point forecasts, quantiles or samples by checkpoint.
Diffusion: Joint path samples conditional on fitted parameters.} \\
\addlinespace
Predictive relevance
& Linear coefficient restrictions or nonlinear conditional-mean or
  conditional-law restrictions, with appropriate tests or estimation theory.
& Controlled input exclusions or structured restrictions can encode
  noncausality; attention weights and selected sparsity alone are not tests. \\
\addlinespace
Structural identification
& Economic restrictions, instruments, or research designs identify the
  shock or intervention separately from predictive fit.
& Predictive flexibility can be combined with identifying assumptions;
  the architecture alone does not identify a policy effect. \\
\bottomrule
\end{tabularx}
\par\endgroup 
\end{table}

\clearpage
\bibliographystyle{plainnat}
\bibliography{JTSA_References}

@inproceedings{wen2023transformers,
  author    = {Wen, Qingsong and Zhou, Tian and Zhang, Chaoli and Chen, Weiqi and Ma, Ziqing and Yan, Junchi and Sun, Liang},
  title     = {Transformers in Time Series: A Survey},
  booktitle = {Proceedings of the Thirty-Second International Joint Conference on Artificial Intelligence (IJCAI)},
  year      = {2023},
  pages     = {6778--6786}
}

@inproceedings{chen2023score,
  author    = {Chen, Minshuo and Huang, Kaixuan and
               Zhao, Tuo and Wang, Mengdi},
  title     = {Score Approximation, Estimation and Distribution
               Recovery of Diffusion Models on Low-Dimensional Data},
  booktitle = {Proceedings of the 40th International Conference
               on Machine Learning},
  series    = {Proceedings of Machine Learning Research},
  volume    = {202},
  pages     = {4672--4712},
  year      = {2023},
  publisher = {PMLR},
  url       = {https://proceedings.mlr.press/v202/chen23o.html}
}

@inproceedings{tang2024adaptivity,
  author    = {Tang, Rong and Yang, Yun},
  title     = {Adaptivity of Diffusion Models to Manifold Structures},
  booktitle = {Proceedings of the 27th International Conference
               on Artificial Intelligence and Statistics},
  series    = {Proceedings of Machine Learning Research},
  volume    = {238},
  pages     = {1648--1656},
  year      = {2024},
  publisher = {PMLR},
  url       = {https://proceedings.mlr.press/v238/tang24a.html}
}

@article{chernozhukov2021lasso,
  author  = {Chernozhukov, Victor and H{\"a}rdle, Wolfgang K.
             and Huang, Chen and Wang, Weining},
  title   = {{LASSO}-Driven Inference in Time and Space},
  journal = {The Annals of Statistics},
  year    = {2021},
  volume  = {49},
  number  = {3},
  pages   = {1702--1735},
  doi     = {10.1214/20-AOS2019}
}

@article{chernozhukov2026spatial,
  author  = {Chernozhukov, Victor and Huang, Chen
             and Wang, Weining},
  title   = {Uniform Inference on High-Dimensional
             Spatial Panel Networks},
  journal = {Journal of Business \& Economic Statistics},
  year    = {2026},
  volume  = {44},
  number  = {1},
  pages   = {348--359},
  doi     = {10.1080/07350015.2025.2530122}
}

@misc{chernozhukov2024arellano,
  author        = {Chernozhukov, Victor
                   and Fern{\'a}ndez-Val, Iv{\'a}n
                   and Huang, Chen and Wang, Weining},
  title         = {{Arellano--Bond LASSO} Estimator
                   for Dynamic Linear Panel Models},
  year          = {2024},
  eprint        = {2402.00584},
  archivePrefix = {arXiv},
  primaryClass  = {econ.EM},
  doi           = {10.48550/arXiv.2402.00584},
  note          = {Working paper}
}

@inproceedings{liu2022nonstationary,
  title={Non-stationary Transformers: Exploring the Stationarity in Time Series Forecasting},
  author={Liu, Yong and Wu, Haixu and Wang, Jianmin and Long, Mingsheng},
  booktitle={Advances in Neural Information Processing Systems},
  year={2022}
}

@article{ansari2024chronos2,
   author={Ansari, Abdul Fatir and Shchur, Oleksandr and K{\"u}ken, Jaris and Auer, Andreas and Han, Boran and Mercado, Pedro and Rangapuram, Syama Sundar and Shen, Huibin and Stella, Lorenzo and Zhang, Xiyuan and others},
  title   = {Chronos-2: From Univariate to Universal Forecasting},
  journal = {arXiv preprint arXiv:2510.15821},
  year    = {2025}
}

@article{li2024automatic,
  author  = {Li, Jie and Fearnhead, Paul and Fryzlewicz, Piotr and Wang, Tengyao},
  title   = {Automatic Change-Point Detection in Time Series via Deep Learning},
  journal = {Journal of the Royal Statistical Society Series B: Statistical Methodology},
  year    = {2024},
  volume  = {86},
  number  = {2},
  pages   = {273--285}
}

@article{sims1980macroeconomics,
  author  = {Sims, Christopher A.},
  title   = {Macroeconomics and Reality},
  journal = {Econometrica},
  year    = {1980},
  volume  = {48},
  number  = {1},
  pages   = {1--48}
}

@article{granger1969investigating,
  author  = {Granger, C. W. J.},
  title   = {Investigating Causal Relations by Econometric Models and Cross-spectral Methods},
  journal = {Econometrica},
  year    = {1969},
  volume  = {37},
  number  = {3},
  pages   = {424--438}
}

@inproceedings{xie2022icl,
  author    = {Xie, Sang Michael and Raghunathan, Aditi and Liang, Percy and Ma, Tengyu},
  title     = {An Explanation of In-Context Learning as Implicit {B}ayesian Inference},
  booktitle = {International Conference on Learning Representations (ICLR)},
  year      = {2022}
}

@inproceedings{sohldickstein2015,
  title     = {Deep Unsupervised Learning using Nonequilibrium Thermodynamics},
  author    = {Sohl-Dickstein, Jascha and Weiss, Eric A. and
               Maheswaranathan, Niru and Ganguli, Surya},
  booktitle = {Proceedings of the 32nd International Conference on Machine
               Learning (ICML)},
  series    = {PMLR},
  volume    = {37},
  pages     = {2256--2265},
  year      = {2015}
}

@article{yan2021scoregrad,
  author  = {Yan, Tijin and Zhang, Hongwei and Zhou, Tong and Zhan, Yufeng and Xia, Yuanqing},
  title   = {{ScoreGrad}: Multivariate Probabilistic Time Series Forecasting with Continuous Energy-based Generative Models},
  journal = {arXiv preprint arXiv:2106.10121},
  year    = {2021}
}

@inproceedings{li2022d3vae,
  author    = {Li, Yan and Lu, Xinjiang and Wang, Yaqing and Dou, Dejing},
  title     = {Generative Time Series Forecasting with Diffusion, Denoise, and Disentanglement},
  booktitle = {Advances in Neural Information Processing Systems (NeurIPS)},
  year      = {2022}
}

@inproceedings{yuan2024diffts,
  author    = {Yuan, Xinyu and Qiao, Yan},
  title     = {{Diffusion-TS}: Interpretable Diffusion for General Time Series Generation},
  booktitle = {International Conference on Learning Representations (ICLR)},
  year      = {2024}
}

@inproceedings{
li2024tmdm,
title={Transformer-Modulated Diffusion Models for Probabilistic Multivariate  Time Series Forecasting},
author={Yuxin Li and Wenchao Chen and Xinyue Hu and Bo Chen and Baolin Sun and Mingyuan Zhou},
booktitle={International Conference on Learning Representations (ICLR)},
year={2024}
}

@inproceedings{song2019score,
  title     = {Generative Modeling by Estimating Gradients of the Data
               Distribution},
  author    = {Song, Yang and Ermon, Stefano},
  booktitle = {Advances in Neural Information Processing Systems (NeurIPS)},
  year      = {2019}
}

@inproceedings{song2023consistency,
  title     = {Consistency Models},
  author    = {Song, Yang and Dhariwal, Prafulla and Chen, Mark and
               Sutskever, Ilya},
  booktitle = {Proceedings of the 40th International Conference on Machine
               Learning (ICML)},
  year      = {2023}
}

@inproceedings{liu2023rectified,
  title     = {Flow Straight and Fast: Learning to Generate and Transfer Data
               with Rectified Flow},
  author    = {Liu, Xingchao and Gong, Chengyue and Liu, Qiang},
  booktitle = {International Conference on Learning Representations (ICLR)},
  year      = {2023}
}

@inproceedings{muller2022pfn,
  author    = {M{\"u}ller, Samuel and Hollmann, Noah and Pineda Arango, Sebastian and Grabocka, Josif and Hutter, Frank},
  title     = {Transformers Can Do {B}ayesian Inference},
  booktitle = {International Conference on Learning Representations (ICLR)},
  year      = {2022}
}

@inproceedings{wies2023learnability,
  author    = {Wies, Noam and Levine, Yoav and Shashua, Amnon},
  title     = {The Learnability of In-Context Learning},
  booktitle = {Advances in Neural Information Processing Systems (NeurIPS)},
  year      = {2023}
}

@article{fryzlewicz2014wbs,
  author  = {Fryzlewicz, Piotr},
  title   = {Wild Binary Segmentation for Multiple Change-Point Detection},
  journal = {The Annals of Statistics},
  year    = {2014},
  volume  = {42},
  number  = {6},
  pages   = {2243--2281}
}

@article{dahlhaus1997fitting,
  author  = {Dahlhaus, Rainer},
  title   = {Fitting Time Series Models to Nonstationary Processes},
  journal = {The Annals of Statistics},
  year    = {1997},
  volume  = {25},
  number  = {1},
  pages   = {1--37}
}

@article{engle1987cointegration,
  author  = {Engle, Robert F. and Granger, C. W. J.},
  title   = {Co-integration and Error Correction: Representation, Estimation, and Testing},
  journal = {Econometrica},
  year    = {1987},
  volume  = {55},
  number  = {2},
  pages   = {251--276}
}

@article{johansen1991estimation,
  author  = {Johansen, S{\o}ren},
  title   = {Estimation and Hypothesis Testing of Cointegration Vectors in Gaussian Vector Autoregressive Models},
  journal = {Econometrica},
  year    = {1991},
  volume  = {59},
  number  = {6},
  pages   = {1551--1580}
}

@book{lutkepohl2005new,
  author    = {L{\"u}tkepohl, Helmut},
  title     = {New Introduction to Multiple Time Series Analysis},
  publisher = {Springer},
  year      = {2005}
}

@book{hamilton1994time,
  author    = {Hamilton, James D.},
  title     = {Time Series Analysis},
  publisher = {Princeton University Press},
  year      = {1994}
}

@inproceedings{vaswani2017attention,
  author    = {Vaswani, Ashish and Shazeer, Noam and Parmar, Niki and Uszkoreit, Jakob and Jones, Llion and Gomez, Aidan N. and Kaiser, Lukasz and Polosukhin, Illia},
  title     = {Attention Is All You Need},
  booktitle = {Advances in Neural Information Processing Systems (NeurIPS)},
  year      = {2017}
}

@inproceedings{zhou2021informer,
  author    = {Zhou, Haoyi and Zhang, Shanghang and Peng, Jieqi and Zhang, Shuai and Li, Jianxin and Xiong, Hui and Zhang, Wancai},
  title     = {Informer: Beyond Efficient Transformer for Long Sequence Time-Series Forecasting},
  booktitle = {Proceedings of the AAAI Conference on Artificial Intelligence},
  year      = {2021}
}

@inproceedings{wu2021autoformer,
  author    = {Wu, Haixu and Xu, Jiehui and Wang, Jianmin and Long, Mingsheng},
  title     = {Autoformer: Decomposition Transformers with Auto-Correlation for Long-Term Series Forecasting},
  booktitle = {Advances in Neural Information Processing Systems (NeurIPS)},
  year      = {2021}
}

@inproceedings{zhou2022fedformer,
  author    = {Zhou, Tian and Ma, Ziqing and Wen, Qingsong and Wang, Xue and Sun, Liang and Jin, Rong},
  title     = {{FEDformer}: Frequency Enhanced Decomposed Transformer for Long-Term Series Forecasting},
  booktitle = {International Conference on Machine Learning (ICML)},
  year      = {2022}
}

@inproceedings{zeng2023transformers,
  author    = {Zeng, Ailing and Chen, Muxi and Zhang, Lei and Xu, Qiang},
  title     = {Are Transformers Effective for Time Series Forecasting?},
  booktitle = {Proceedings of the AAAI Conference on Artificial Intelligence},
  year      = {2023}
}

@inproceedings{nie2023patchtst,
  author    = {Nie, Yuqi and Nguyen, Nam H. and Sinthong, Phanwadee and Kalagnanam, Jayant},
  title     = {A Time Series Is Worth 64 Words: Long-Term Forecasting with Transformers},
  booktitle = {International Conference on Learning Representations (ICLR)},
  year      = {2023}
}

@inproceedings{liu2024itransformer,
  author    = {Liu, Yong and Hu, Tengge and Zhang, Haoran and Wu, Haixu and Wang, Shiyu and Ma, Lintao and Long, Mingsheng},
  title     = {{iTransformer}: Inverted Transformers Are Effective for Time Series Forecasting},
  booktitle = {International Conference on Learning Representations (ICLR)},
  year      = {2024}
}

@article{salinas2020deepar,
  author  = {Salinas, David and Flunkert, Valentin and Gasthaus, Jan and Januschowski, Tim},
  title   = {{DeepAR}: Probabilistic Forecasting with Autoregressive Recurrent Networks},
  journal = {International Journal of Forecasting},
  year    = {2020},
  volume  = {36},
  number  = {3},
  pages   = {1181--1191}
}

@inproceedings{oreshkin2020nbeats,
  author    = {Oreshkin, Boris N. and Carpov, Dmitri and Chapados, Nicolas and Bengio, Yoshua},
  title     = {{N-BEATS}: Neural Basis Expansion Analysis for Interpretable Time Series Forecasting},
  booktitle = {International Conference on Learning Representations (ICLR)},
  year      = {2020}
}

@article{
ansari2024chronos,
title={Chronos: Learning the Language of Time Series},
author={Abdul Fatir Ansari and Lorenzo Stella and Ali Caner Turkmen and Xiyuan Zhang and Pedro Mercado and Huibin Shen and Oleksandr Shchur and Syama Sundar Rangapuram and Sebastian Pineda Arango and Shubham Kapoor and Jasper Zschiegner and Danielle C. Maddix and Hao Wang and Michael W. Mahoney and Kari Torkkola and Andrew Gordon Wilson and Michael Bohlke-Schneider and Bernie Wang},
journal={Transactions on Machine Learning Research},
issn={2835-8856},
year={2024}
}

@article{hyndman2008automatic,
  author  = {Hyndman, Rob J. and Khandakar, Yeasmin},
  title   = {Automatic Time Series Forecasting: The forecast Package for {R}},
  journal = {Journal of Statistical Software},
  year    = {2008},
  volume  = {27},
  number  = {3},
  pages   = {1--22}
}

@article{liu2025moirai2,
  author  = {Liu, Chenghao and Aksu, Taha and Liu, Juncheng and Liu, Xu and Yan, Hanshu and Pham, Quang and Savarese, Silvio and Sahoo, Doyen and Xiong, Caiming and Li, Junnan},
  title   = {Moirai 2.0: When Less Is More for Time Series Forecasting},
  journal = {arXiv preprint arXiv:2511.11698},
  year    = {2025}
}

@inproceedings{das2024timesfm,
  author    = {Das, Abhimanyu and Kong, Weihao and Sen, Rajat and Zhou, Yichen},
  title     = {A Decoder-Only Foundation Model for Time-Series Forecasting},
  booktitle = {International Conference on Machine Learning (ICML)},
  year      = {2024}
}

@inproceedings{woo2024moirai,
  author    = {Woo, Gerald and Liu, Chenghao and Kumar, Akshat and Xiong, Caiming and Savarese, Silvio and Sahoo, Doyen},
  title     = {Unified Training of Universal Time Series Forecasting Transformers},
  booktitle = {International Conference on Machine Learning (ICML)},
  year      = {2024}
}

@article{garza2023timegpt,
  author  = {Garza, Azul and Challu, Cristian and Mergenthaler-Canseco, Max},
  title   = {{TimeGPT-1}},
  journal = {arXiv preprint arXiv:2310.03589},
  year    = {2023}
}

@article{rasul2023lagllama,
  author  = {Rasul, Kashif and Ashok, Arjun and Williams, Andrew Robert and Ghonia, Hena
and Bhagwatkar, Rishika and Khorasani, Arian and Darvishi Bayazi, Mohammad Javad
and Adamopoulos, George and Riachi, Roland and Hassen, Nadhir and Bilo{\v{s}}, Marin
and Garg, Sahil and Schneider, Anderson and Chapados, Nicolas and Drouin, Alexandre
and Zantedeschi, Valentina and Nevmyvaka, Yuriy and Rish, Irina},
  title   = {Lag-Llama: Towards Foundation Models for Probabilistic Time Series Forecasting},
  journal = {arXiv preprint arXiv:2310.08278},
  year    = {2023}
}

@inproceedings{shi2024timemoe,
  author    = {Shi, Xiaoming and Wang, Shiyu and Nie, Yuqi and Li, Dianqi and Ye, Zhou and Wen, Qingsong and Jin, Ming},
  title     = {{Time-MoE}: Billion-Scale Time Series Foundation Models with Mixture of Experts},
  booktitle = {International Conference on Learning Representations (ICLR)},
  year      = {2025}
}

@article{aksu2024gifteval,
  author  = {Aksu, Taha and Woo, Gerald and Liu, Juncheng and Liu, Xu and Liu, Chenghao and Savarese, Silvio and Xiong, Caiming and Sahoo, Doyen},
  title   = {{GIFT-Eval}: A Benchmark for General Time Series Forecasting Model Evaluation},
  journal = {arXiv preprint arXiv:2410.10393},
  year    = {2024}
}

@inproceedings{ho2020denoising,
  author    = {Ho, Jonathan and Jain, Ajay and Abbeel, Pieter},
  title     = {Denoising Diffusion Probabilistic Models},
  booktitle = {Advances in Neural Information Processing Systems (NeurIPS)},
  year      = {2020}
}

@inproceedings{song2021scorebased,
  author    = {Song, Yang and Sohl-Dickstein, Jascha and Kingma, Diederik P. and Kumar, Abhishek and Ermon, Stefano and Poole, Ben},
  title     = {Score-Based Generative Modeling through Stochastic Differential Equations},
  booktitle = {International Conference on Learning Representations (ICLR)},
  year      = {2021}
}

@inproceedings{song2021ddim,
  author    = {Song, Jiaming and Meng, Chenlin and Ermon, Stefano},
  title     = {Denoising Diffusion Implicit Models},
  booktitle = {International Conference on Learning Representations (ICLR)},
  year      = {2021}
}

@inproceedings{lee2024ant,
  title={ANT: Adaptive Noise Schedule for Time Series Diffusion Models},
  author={Lee, Seunghan and Lee, Kibok and Park, Taeyoung},
  booktitle={Advances in Neural Information Processing Systems (NeurIPS)},
  year={2024}
}

@article{GRANGER1974111,
title = {Spurious regressions in econometrics},
journal = {Journal of Econometrics},
volume = {2},
number = {2},
pages = {111-120},
year = {1974},
issn = {0304-4076},
author = {C.W.J. Granger and P. Newbold}
}

@article{challu2023nhits,
  title={{NHITS}: Neural Hierarchical Interpolation for Time Series Forecasting},
author={Challu, Cristian and Olivares, Kin G. and Oreshkin, Boris N. and Garza Ramirez, Federico and Mergenthaler Canseco, Max and Dubrawski, Artur},
  journal={Proceedings of the AAAI Conference on Artificial Intelligence},
  volume={37},
  number={6},
  pages={6989--6997},
  year={2023}
}

@article{marcellino2006comparison,
  author  = {Marcellino, Massimiliano and Stock, James H. and Watson, Mark W.},
  title   = {A Comparison of Direct and Iterated Multistep {AR} Methods for Forecasting Macroeconomic Time Series},
  journal = {Journal of Econometrics},
  year    = {2006},
  volume  = {135},
  number  = {1--2},
  pages   = {499--526}
}

@inproceedings{lipman2023flow,
  author    = {Lipman, Yaron and Chen, Ricky T. Q. and Ben-Hamu, Heli and Nickel, Maximilian and Le, Matt},
  title     = {Flow Matching for Generative Modeling},
  booktitle = {International Conference on Learning Representations (ICLR)},
  year      = {2023}
}

@inproceedings{rasul2021timegrad,
  author    = {Rasul, Kashif and Seward, Calvin and Schuster, Ingmar and Vollgraf, Roland},
  title     = {Autoregressive Denoising Diffusion Models for Multivariate Probabilistic Time Series Forecasting},
  booktitle = {International Conference on Machine Learning (ICML)},
  year      = {2021}
}

@inproceedings{tashiro2021csdi,
  author    = {Tashiro, Yusuke and Song, Jiaming and Song, Yang and Ermon, Stefano},
  title     = {{CSDI}: Conditional Score-based Diffusion Models for Probabilistic Time Series Imputation},
  booktitle = {Advances in Neural Information Processing Systems (NeurIPS)},
  year      = {2021}
}

@article{alcaraz2023sssd,
  author  = {Alcaraz, Juan Miguel Lopez and Strodthoff, Nils},
  title   = {Diffusion-based Time Series Imputation and Forecasting with Structured State Space Models},
  journal = {Transactions on Machine Learning Research (TMLR)},
  year    = {2023}
}

@inproceedings{shen2023timediff,
  author    = {Shen, Lifeng and Kwok, James T.},
  title     = {Non-autoregressive Conditional Diffusion Models for Time Series Prediction},
  booktitle = {International Conference on Machine Learning (ICML)},
  year      = {2023}
}

@inproceedings{kollovieh2023tsdiff,
  author    = {Kollovieh, Marcel and Ansari, Abdul Fatir and Bohlke-Schneider, Michael and Zschiegner, Jasper and Wang, Hao and Wang, Yuyang},
  title     = {Predict, Refine, Synthesize: Self-Guiding Diffusion Models for Probabilistic Time Series Forecasting},
  booktitle = {Advances in Neural Information Processing Systems (NeurIPS)},
  year      = {2023}
}

@inproceedings{fan2024mgtsd,
  author    = {Fan, Xinyao and Wu, Yueying and Xu, Chang and Huang, Yuhao and Liu, Weiqing and Bian, Jiang},
  title     = {{MG-TSD}: Multi-Granularity Time Series Diffusion Models with Guided Learning Process},
  booktitle = {International Conference on Learning Representations (ICLR)},
  year      = {2024}
}

@inproceedings{shen2024mrdiff,
  author    = {Shen, Lifeng and Chen, Weiyu and Kwok, James T.},
  title     = {Multi-Resolution Diffusion Models for Time Series Forecasting},
  booktitle = {International Conference on Learning Representations (ICLR)},
  year      = {2024}
}

@inproceedings{gu2022s4,
  author    = {Gu, Albert and Goel, Karan and R{\'e}, Christopher},
  title     = {Efficiently Modeling Long Sequences with Structured State Spaces},
  booktitle = {International Conference on Learning Representations (ICLR)},
  year      = {2022}
}

@article{yang2024survey,
  author  = {Yang, Yiyuan and Jin, Ming and Wen, Haomin and Zhang, Chaoli and Liang, Yuxuan and Ma, Lintao and Wang, Yi and Liu, Chenghao and Yang, Bin and Xu, Zenglin and Bian, Jiang and Pan, Shirui and Wen, Qingsong},
  title   = {A Survey on Diffusion Models for Time Series and Spatio-Temporal Data},
  journal = {arXiv preprint arXiv:2404.18886},
  year    = {2024}
}

@article{meijer2024rise,
  author  = {Meijer, Caspar and Chen, Lydia Y.},
  title   = {The Rise of Diffusion Models in Time-Series Forecasting},
  journal = {arXiv preprint arXiv:2401.03006},
  year    = {2024}
}

@article{gneiting2007strictly,
  author  = {Gneiting, Tilmann and Raftery, Adrian E.},
  title   = {Strictly Proper Scoring Rules, Prediction, and Estimation},
  journal = {Journal of the American Statistical Association},
  year    = {2007},
  volume  = {102},
  number  = {477},
  pages   = {359--378}
}

@inproceedings{kim2022revin,
  author    = {Kim, Taesung and Kim, Jinhee and Tae, Yunwon and Park, Cheonbok and Choi, Jang-Ho and Choo, Jaegul},
  title     = {Reversible Instance Normalization for Accurate Time-Series Forecasting against Distribution Shift},
  booktitle = {International Conference on Learning Representations (ICLR)},
  year      = {2022}
}

@article{litterman1986forecasting,
  author  = {Litterman, Robert B.},
  title   = {Forecasting with {B}ayesian Vector Autoregressions---Five Years of Experience},
  journal = {Journal of Business \& Economic Statistics},
  year    = {1986},
  volume  = {4},
  number  = {1},
  pages   = {25--38}
}

@article{banbura2010large,
  author  = {Ba\'nbura, Marta and Giannone, Domenico and Reichlin, Lucrezia},
  title   = {Large {B}ayesian Vector Auto Regressions},
  journal = {Journal of Applied Econometrics},
  year    = {2010},
  volume  = {25},
  number  = {1},
  pages   = {71--92}
}

@article{stock2002forecasting,
  author  = {Stock, James H. and Watson, Mark W.},
  title   = {Forecasting Using Principal Components from a Large Number of Predictors},
  journal = {Journal of the American Statistical Association},
  year    = {2002},
  volume  = {97},
  number  = {460},
  pages   = {1167--1179}
}

@article{baing2002,
  author  = {Bai, Jushan and Ng, Serena},
  title   = {Determining the Number of Factors in Approximate Factor Models},
  journal = {Econometrica},
  year    = {2002},
  volume  = {70},
  number  = {1},
  pages   = {191--221}
}

@article{basu2015,
  author  = {Basu, Sumanta and Michailidis, George},
  title   = {Regularized Estimation in Sparse High-Dimensional Time Series Models},
  journal = {The Annals of Statistics},
  year    = {2015},
  volume  = {43},
  number  = {4},
  pages   = {1535--1567}
}

@article{kock2015,
  author  = {Kock, Anders Bredahl and Callot, Laurent},
  title   = {Oracle Inequalities for High Dimensional Vector Autoregressions},
  journal = {Journal of Econometrics},
  year    = {2015},
  volume  = {186},
  number  = {2},
  pages   = {325--344}
}

@article{dickey1979,
  author  = {Dickey, David A. and Fuller, Wayne A.},
  title   = {Distribution of the Estimators for Autoregressive Time Series with a Unit Root},
  journal = {Journal of the American Statistical Association},
  year    = {1979},
  volume  = {74},
  number  = {366},
  pages   = {427--431}
}

@article{phillips1988,
  author  = {Phillips, Peter C. B. and Perron, Pierre},
  title   = {Testing for a Unit Root in Time Series Regression},
  journal = {Biometrika},
  year    = {1988},
  volume  = {75},
  number  = {2},
  pages   = {335--346}
}

@article{blanchard1989,
  author  = {Blanchard, Olivier Jean and Quah, Danny},
  title   = {The Dynamic Effects of Aggregate Demand and Supply Disturbances},
  journal = {The American Economic Review},
  year    = {1989},
  volume  = {79},
  number  = {4},
  pages   = {655--673}
}

@book{kilian2017,
  author    = {Kilian, Lutz and L\"utkepohl, Helmut},
  title     = {Structural Vector Autoregressive Analysis},
  publisher = {Cambridge University Press},
  year      = {2017}
}

@book{johansen1995,
  author    = {Johansen, S{\o}ren},
  title     = {Likelihood-Based Inference in Cointegrated Vector Autoregressive Models},
  publisher = {Oxford University Press},
  year      = {1995}
}

@article{bai1998breaks,
 author = {Bai, Jushan and Perron, Pierre},
 title = {Estimating and Testing Linear Models with Multiple Structural Changes},
 journal = {Econometrica}, year = {1998}, volume = {66}, number = {1}, pages = {47--78},
 doi = {10.2307/2998540}, url = {https://doi.org/10.2307/2998540}
}

@article{breiman2001forests,
 author = {Breiman, Leo}, title = {Random Forests}, journal = {Machine Learning},
 year = {2001}, volume = {45}, number = {1}, pages = {5--32},
 doi = {10.1023/A:1010933404324}, url = {https://doi.org/10.1023/A:1010933404324}
}

@article{friedman2001boosting,
 author = {Friedman, Jerome H.}, title = {Greedy Function Approximation: A Gradient Boosting Machine},
 journal = {The Annals of Statistics}, year = {2001}, volume = {29}, number = {5}, pages = {1189--1232},
 doi = {10.1214/aos/1013203451}, url = {https://doi.org/10.1214/aos/1013203451}
}

@article{goulet2022ml,
 author = {Goulet Coulombe, Philippe and Leroux, Maxime and Stevanovic, Dalibor and Surprenant, St{\'e}phane},
 title = {How Is Machine Learning Useful for Macroeconomic Forecasting?},
 journal = {Journal of Applied Econometrics}, year = {2022}, volume = {37}, number = {5}, pages = {920--964},
 doi = {10.1002/jae.2910}, url = {https://doi.org/10.1002/jae.2910}
}

@article{hansen2002threshold,
 author = {Hansen, Bruce E. and Seo, Byeongseon},
 title = {Testing for Two-Regime Threshold Cointegration in Vector Error-Correction Models},
 journal = {Journal of Econometrics}, year = {2002}, volume = {110}, number = {2}, pages = {293--318},
 doi = {10.1016/S0304-4076(02)00097-0}, url = {https://doi.org/10.1016/S0304-4076(02)00097-0}
}

@article{januschowski2022trees,
 author = {Januschowski, Tim and Wang, Yuyang and Torkkola, Kari and Erkkil{\"a}, Timo and Hasson, Hilaf and Gasthaus, Jan},
 title = {Forecasting with Trees}, journal = {International Journal of Forecasting},
 year = {2022}, volume = {38}, number = {4}, pages = {1473--1481},
 doi = {10.1016/j.ijforecast.2021.10.004}, url = {https://doi.org/10.1016/j.ijforecast.2021.10.004}
}

@book{krolzig1997markov,
 author = {Krolzig, Hans-Martin},
 title = {Markov-Switching Vector Autoregressions: Modelling, Statistical Inference, and Application to Business Cycle Analysis},
 publisher = {Springer}, address = {Berlin, Heidelberg},
 series = {Lecture Notes in Economics and Mathematical Systems}, volume = {454}, year = {1997},
 doi = {10.1007/978-3-642-51684-9}, url = {https://doi.org/10.1007/978-3-642-51684-9}
}

@article{masini2023ml,
 author = {Masini, Ricardo P. and Medeiros, Marcelo C. and Mendes, Eduardo F.},
 title = {Machine Learning Advances for Time Series Forecasting},
 journal = {Journal of Economic Surveys}, year = {2023}, volume = {37}, number = {1}, pages = {76--111},
 doi = {10.1111/joes.12429}, url = {https://doi.org/10.1111/joes.12429}
}

@article{primiceri2005tvp,
 author = {Primiceri, Giorgio E.}, title = {Time Varying Structural Vector Autoregressions and Monetary Policy},
 journal = {The Review of Economic Studies}, year = {2005}, volume = {72}, number = {3}, pages = {821--852},
 doi = {10.1111/j.1467-937X.2005.00353.x}, url = {https://doi.org/10.1111/j.1467-937X.2005.00353.x}
}

@article{tibshirani1996lasso,
 author = {Tibshirani, Robert}, title = {Regression Shrinkage and Selection via the Lasso},
 journal = {Journal of the Royal Statistical Society: Series B (Methodological)},
 year = {1996}, volume = {58}, number = {1}, pages = {267--288},
 doi = {10.1111/j.2517-6161.1996.tb02080.x}, url = {https://doi.org/10.1111/j.2517-6161.1996.tb02080.x}
}

@article{tsay1998threshold,
 author = {Tsay, Ruey S.}, title = {Testing and Modeling Multivariate Threshold Models},
 journal = {Journal of the American Statistical Association},
 year = {1998}, volume = {93}, number = {443}, pages = {1188--1202},
 doi = {10.1080/01621459.1998.10473779}, url = {https://doi.org/10.1080/01621459.1998.10473779}
}

@article{zou2005elastic,
 author = {Zou, Hui and Hastie, Trevor}, title = {Regularization and Variable Selection via the Elastic Net},
 journal = {Journal of the Royal Statistical Society: Series B (Statistical Methodology)},
 year = {2005}, volume = {67}, number = {2}, pages = {301--320},
 doi = {10.1111/j.1467-9868.2005.00503.x}, url = {https://doi.org/10.1111/j.1467-9868.2005.00503.x}
}

@inproceedings{kim2021vilt,
  author = {Kim, Wonjae and Son, Bokyung and Kim, Ildoo},
  title = {{ViLT}: Vision-and-Language Transformer Without Convolution or Region Supervision},
  booktitle = {Proceedings of the 38th International Conference on Machine Learning},
  series = {Proceedings of Machine Learning Research},
  volume = {139}, pages = {5583--5594}, year = {2021},
  publisher = {PMLR},
  url = {https://proceedings.mlr.press/v139/kim21k.html}
}

@techreport{antolin2026attention,
 author = {Antol{\'i}n-D{\'i}az, Juan and Rubio-Ram{\'i}rez, Juan F.},
 title = {Attention-Based Vector Autoregressions},
 institution = {SSRN},
 number = {6479559},
 year = {2026},
 note = {Working paper, dated 26 March 2026; posted 17 April 2026},
 doi = {10.2139/ssrn.6479559},
 url = {https://papers.ssrn.com/sol3/papers.cfm?abstract_id=6479559}
}

@article{koop1996impulse,
  author = {Koop, Gary and Pesaran, M. Hashem and Potter, Simon M.},
  title = {Impulse Response Analysis in Nonlinear Multivariate Models},
  journal = {Journal of Econometrics},
  year = {1996}, volume = {74}, number = {1}, pages = {119--147},
  doi = {10.1016/0304-4076(95)01753-4},
  url = {https://doi.org/10.1016/0304-4076(95)01753-4}
}

@article{tank2022neural,
  author = {Tank, Alex and Covert, Ian and Foti, Nicholas and Shojaie, Ali and Fox, Emily B.},
  title = {Neural {Granger} Causality},
  journal = {IEEE Transactions on Pattern Analysis and Machine Intelligence},
  year = {2022}, volume = {44}, number = {8}, pages = {4267--4279},
  doi = {10.1109/TPAMI.2021.3065601},
  url = {https://doi.org/10.1109/TPAMI.2021.3065601}
}

\end{document}